\documentclass[aps,prx,twocolumn,superscriptaddress,showpacs,floatfix]{revtex4-2}
\usepackage{color}
\usepackage{mathtools}
\usepackage[c]{esvect}
\usepackage{bm}        
\usepackage{amssymb}   
\usepackage{amssymb, amsthm, float, graphicx, amsfonts, dsfont}
\usepackage{amsmath} 

\usepackage[colorlinks=true,citecolor=blue,linkcolor=magenta, breaklinks=true]{hyperref}
\usepackage{graphicx}
\usepackage{braket}
\usepackage{natbib}
\usepackage [latin1]{inputenc}
\usepackage{siunitx}
\usepackage{blkarray, bigstrut}
\usepackage[dvipsnames]{xcolor}
\usepackage{comment}
\usepackage{physics}
\usepackage{tikz}

\newcolumntype{P}[1]{>{\centering\arraybackslash}p{#1}}
\begin{document}
\title{Emergent scalar-chirality \& colossal transverse-magnetoresponse in strongly correlated nodal-line half-metal.}

\author{Jyotirmoy Sau}
\thanks{\textcolor{magenta}{These authors contributed equally to this work.}}
\affiliation{Department of Condensed Matter and Materials Physics,
S. N. Bose National Centre for Basic Sciences, JD Block, Sector III, Salt Lake, Kolkata 700106, India}

\author{Sourav Chakraborty}
\thanks{\textcolor{magenta}{These authors contributed equally to this work.}}
\affiliation{Department of Condensed Matter and Materials Physics,
S. N. Bose National Centre for Basic Sciences, JD Block, Sector III, Salt Lake, Kolkata 700106, India}

\author{Sourabh Saha}
\affiliation{Department of Condensed Matter and Materials Physics,
S. N. Bose National Centre for Basic Sciences, JD Block, Sector III, Salt Lake, Kolkata 700106, India}

\author{Kalpataru Pradhan}
\email{kalpataru.pradhan@saha.ac.in}
\affiliation{Theory Division, Saha Institute of Nuclear Physics,
  A CI of Homi Bhabha National Institute, Kolkata-700064, India}

\author{Anamitra Mukherjee}
\email{anamitra@niser.ac.in}
\affiliation{School of Physical Sciences, National Institute of Science Education and Research,
a CI of Homi Bhabha National Institute, Jatni 752050, India}

\author{Manoranjan Kumar}
\email{manoranjan.kumar@bose.res.in}
\affiliation{Department of Condensed Matter and Materials Physics,
S. N. Bose National Centre for Basic Sciences, JD Block, Sector III, Salt Lake, Kolkata 700106, India}

\begin{abstract}
Understanding the interplay of strong correlation and temperature in nodal-line semimetals can offer novel ways to control spin currents. Here we consider the 3d-5d double-perovskite Ba$_{2}$CoWO$_{6}$, which features mirror-symmetry-protected nodal-lines, strong Co-site interactions, and spin-orbit coupling (SOC) at W sites. Our first principles and exact diagonalization results reveal a half-metallic ground state with high-spin Co and topologically non-trivial bands. We demonstrate that SOC gaps out nodal points, causes band-inversion and generates anomalous Hall response. A semi-classical Monte Carlo finite-temperature simulation of five-orbital Hubbard model uncovers an emergent Co-spin scalar chirality and colossal positive transverse-magnetoresponse. We predict the temperature and magnetic field scales for the tunability of scalar-chirality and magnetoresponse.
\end{abstract}

\maketitle
\textit{Introduction:--- } Crystalline-symmetry-protected topological insulators \cite{cti-fu} has ushered a new paradigm of research on the coexistence of broken time-reversal and inversion symmetry along with band topology \cite{topo-metal-review-th, topo-metal-review-exp}. Dirac \cite{dirac-th-1,dirac-th-2,dirac-th-3}, nodal-line\cite{semi-metal-dft-review-2} and Weyl \cite{Weyl-th-1,Weyl-th-2} semi-metal have been theoretically proposed and experimentally realized \cite{dirac-exp-1, Weyl-exp-1,Weyl-exp-2}. In particular, half-metals with fully spin-polarized conduction bands bring together correlation effects and crystalline symmetry-induced band-topology, two vital ingredients for realizing magnetic-topological metals. While the magnetic order can reduce the crystalline symmetry, it allows for the survival of some nodal lines, classified by generalized Chern numbers \cite{mirror-chern-1,mirror-chern-2}. In the presence of spin-orbit coupling (SOC), these remnant nodal lines can be gapped out along with band-inversion at nodal points \cite{mirror}. When the nodal points are close to the Fermi energy, the band-inversion leads to anomalous Hall conductivity (AHC) in spin-polarized bands. The transverse spin-current generated as a consequence has potential applications in low-energy electronics \cite{application-hall}, spintronics \cite{application-spin}, and quantum computation  \cite{application-quant}. Thus, understanding the interplay of correlation effects induced magnetism, crystalline symmetry protection, and spin-orbit effects is vital. 

Double perovskites (DP) half-metals offer a unique platform where crystalline symmetry, strong Coulomb correlation, and SOC can compete. Unlike simple perovskites, in DP, the interaction and SOC effects occur at distinct atomic sites. In many DP of the form, A$_2$BB$^\prime$O$_6$ half-metallicity typically originates from correlation-induced local moments at B(=3d) atom and carriers delocalizing in the  B-B$^\prime$(=4d/5d) structure \cite{half-metal-rmp,half-metal-review-1}. The latter B$^\prime$(=4d/5d) is the source of SOC. DP's are also known to host mirror planes and are natural candidates for crystalline-protected bands. From an experimental standpoint, the considerable charge transfer energy between 3d and 5d atoms allows anti-site defect-free growth \cite{antisite-dis-old,antisite-dis-1}. However, theoretical investigations of these important systems have been limited to first-principles \cite{semi-metal-dft-review-1, semi-metal-dft-review-2} and zero-temperature mean-field studies assuming classical spin-moments at B-site  \cite{bfro-bfwo-prl-cook,bfr-prb-cook}. While these studies have added valuable insights, both first-principles and mean-field approach studies miss out on strong correlation effects. Further, classical spin treatment cannot capture the formation and evolution of B-site local magnetic moment with temperature.
Thus, the question of magnetization at the B site evolves when electrons traverse in a topological band, the impact of topology on their magnetic order and transport, and its temperature and magnetic field response has largely remained open. The dearth of results is due to the enormous complexity of modeling multi-orbital models and the associated computational demands. Nevertheless, addressing these questions is a fundamental theoretical challenge in materials theory and is paramount for future technological applications. In this letter, we combine first-principles study, exact diagonalization, and a semi-classical Monte-Carlo approach to address these questions.

We consider the double perovskite Ba$_{2}$CoWO$_{6}$ (BCWO) containing 3d and 5d transition metal atoms, where the former introduces a strong local correlation effect, and the latter provides a spin-orbit coupling. Co$^{2+}$ is nominally in a $d^7$ state with two $t_{2g}$ orbitals doubly occupied, the third $t_{2g}$ level, and two $e_{g}$ orbitals singly occupied, while W is in a $d^0$ state. 
Through first-principles calculations, we demonstrate a half-metallic ground state in BCWO with a Co high-spin state providing localized moments (majority spins) and delocalized minority carriers. Our analysis shows that, without spin-orbit coupling (SOC), the band structure supports mirror-symmetry-protected nodal lines, two of which are gapped out due to magnetic order. Introducing SOC on W gaps out the remaining nodal line induces band inversion at the nodal points and generates AHC due to Berry curvature effects.
Next, we construct a down-folded five-orbital Hubbard model with full Kanamori parmeterization\cite{kanamori}, confirm the ground state using exact diagonalization, and track the temperature evolution of magnetic order and charge transport with a semi-classical Monte Carlo approach. We uncover a remarkable manifestation of topology in the Co-site magnetic order as an emergent non-coplanar spin texture. We show that the non-coplanarity is generated without adding Dzyaloshinskii-Moriya interaction (DMI) and is consistent with non-zero Berry curvature and transverse conductivity. To demonstrate the topological origin of the non-coplanarity, we show that an external magnetic that suppresses the non-coplanarity also suppresses the AHC while minimally affecting the \textit{longitudinal} response. The combined temperature and magnetic field investigation of transport leads to a colossal \textit{transverse} magnet response. 
We predict the temperature scale for the survival of AHC and the magneto-response systematics, offering an experimental knob for controlling AHC.

\begin{figure}[t]
\centering
\includegraphics[width=1.05\linewidth]{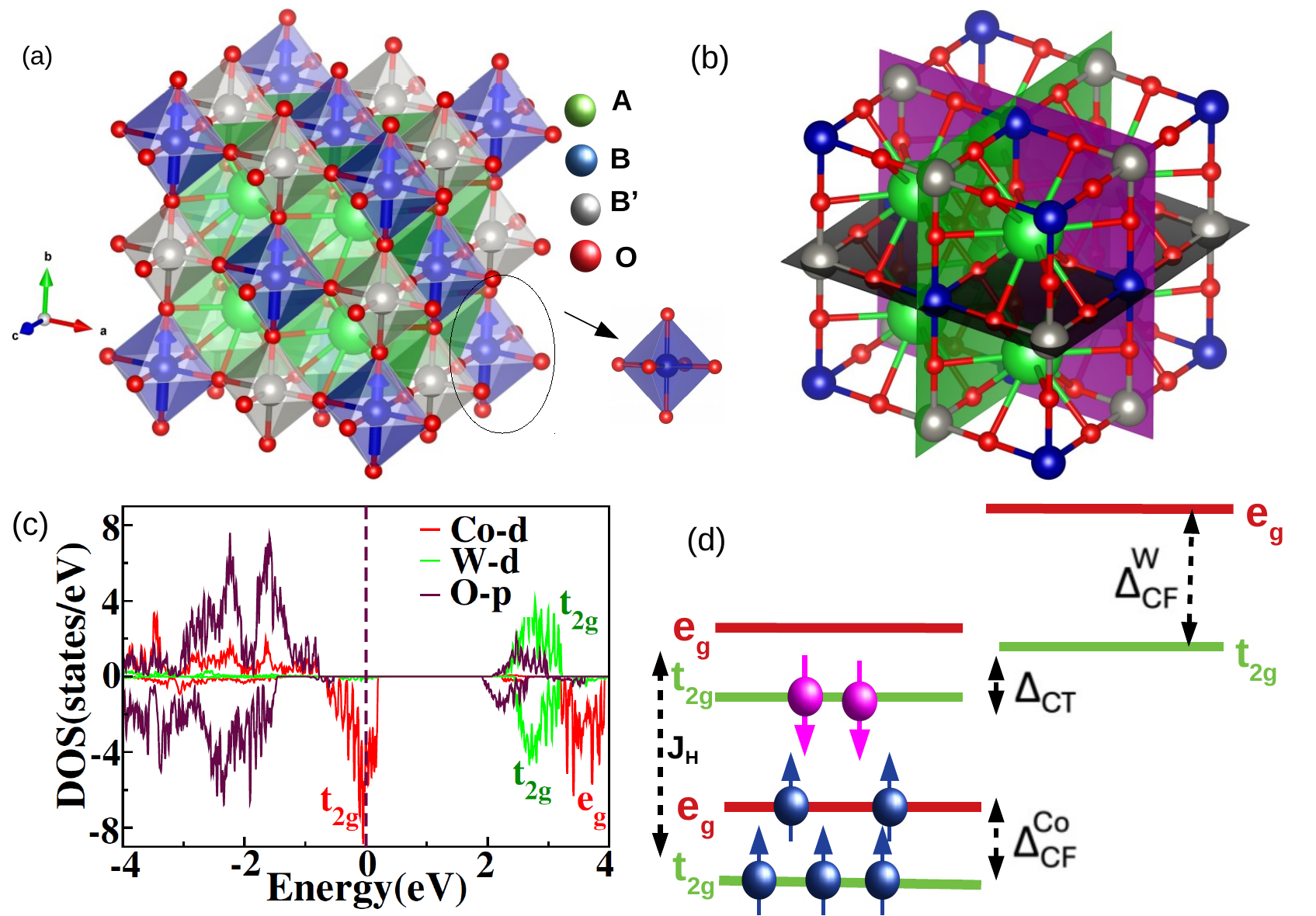}
\caption{\textbf{Crystal \& electronic structure of Ba$_{2}$CoW$O_{6}$:} (a) shows the BCWO crystal structure where green, blue, gray, and red balls represent Ba, Co, W and O respectively. Blue arrows indicate direction of 
along [001]. (b) shows the three mirror planes (M$_x$, M$_y$, M$_z$). (c) spin-polarized orbital-resolved density of states of Co-$d$,O-$p$ and W-$d$ orbitals from density functional theory (DFT) without spin-orbit coupling, where dashed line indicates Fermi energy. (d) Schematic electronic level diagram based on the SOC-free first principles results, where $\Delta_{CF}^{Co}$ and $\Delta_{CF}^{W}$ are crystal field splitting between $t_{2g}$and $e_g$ orbitals, $J_H$ is the Hund's coupling and $\Delta_{CT}$ is the Co-W charge transfer energy.}
\label{fig1}
\end{figure}

\textit{Electronic Structure:--- } We compute the band structure of BCWO, which is a face-centered cubic crystal with the Co and W atoms surrounded by an octahedral cage of oxygen as shown in Fig. \ref{fig1}(a). Details are provided in \textit{Sec. IA of Supplemental Materials}\cite{suppmat}. The space group of the crystal is \textit{Fm$\bar{3}$m} (space group no. 225), which possesses O$_{h}^5$ octahedral symmetry and a lattice parameter of a = 8.210 \r{A}\cite{bwco-hm-dft-old}. The structure exhibits three mirror planes (M$_x$, M$_y$, M$_z$), illustrated in Fig.\ref{fig1}(b). The spin and orbital projected density of states (DOS) in Fig.\ref{fig1}(c) without SOC shows that the metallic nature originates from the Co and W $t_{2g}$ $\downarrow$-spin (\textit{minority}) channel electrons with dominant (small) spectral weight from Co (W) at $E_F$. The charge gap in the  $\uparrow$  spin (\textit{majority}) spin channel is 2.5 eV. The Bader charge analysis indicates that Co2$^+$ (3d$^7$) is in the high spin state S=3/2 and W6$^+$ (5d$^0$) has S $=$ 0, giving a net magnetic moment of  3.0 $\mu_B$/f.u \cite{bwco-hm-dft-old}. The corresponding electronic level diagram in Fig.\ref{fig1}(d). From the first-principles analysis, we extract the Co crystal field splitting $\Delta_{CF}^{Co}$ to be 1.2 eV, the W crystal field splitting $\Delta_{CF}^{W}$ to be 8eV to be and a Co-W charge transfer energy $\Delta_{CT}$ to be 2.5 eV. The high spin state of Co arises due to large Hund's coupling ($J_H$). The hybridization between Co $t_{2g}\downarrow $ and W $t_{2g}$ stabilizes the half-metallic ground state as also seen from Wannierization calculations (see \textit{Supplemental Material }\cite{suppmat} \textit{Sec IA}).

\begin{figure}[t]
\centering
\includegraphics[width=1.0\linewidth]{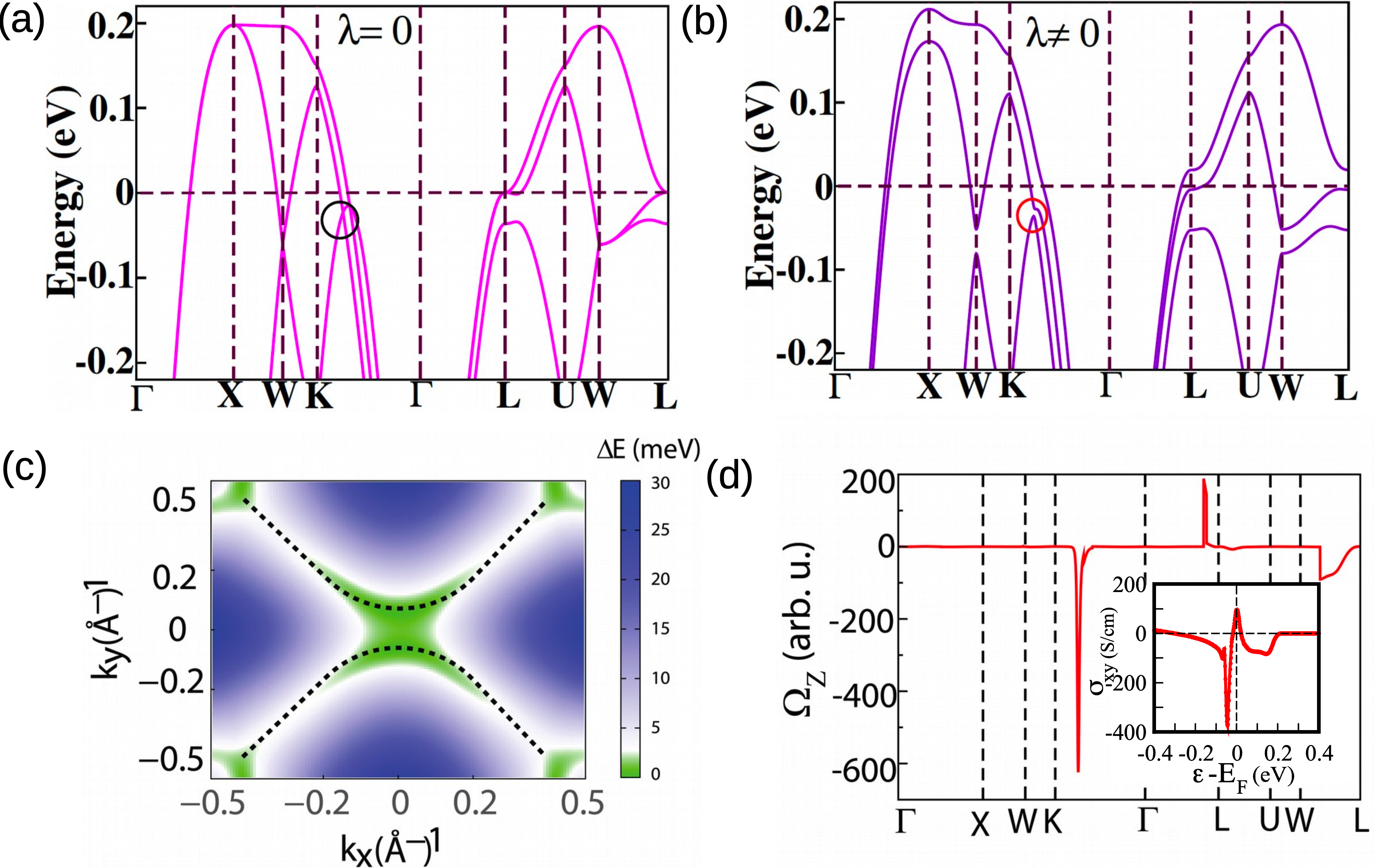}
\caption{\textbf{Band topology \& Berry curvature:}  The band structure of BCWO without (a) SOC and with SOC (b) along high symmetry directions. The nodal points in (a), marked by the black circle, are gapped out in (b), indicated by red circles. (c) The energy gap between two intersecting bands \textit{in the vicinity of} $E_F$ on the k$_x$-k$_y$ plane. The nodal lines for zero SOC (thin black dashed line) are gapped out by SOC. (d) The Berry curvature along the high-symmetry directions. Inset shows the AHC as a function of filling corresponding to (b).}
\label{fig4}
\end{figure}

\textit{Crystalline symmetries \& band topology:--- } We show the spin polarized ($\uparrow$-blue, $\downarrow$-magenta) band structure along the the high-symmetry direction in Fig.\ref{fig4}(a).  A linear band crossing point (circle) between Co d$_{xz}$ and d$_{yz}$ orbitals (of the $\downarrow$-spin bands) is observed along the $K$ to $\Gamma$ direction, just below the Fermi energy ($E_F$). We provide an expanded view of the crossing in Sec. IB of the\textit{ Supplemental Material}\cite{suppmat}. The crossing of the singly degenerate minority band strongly indicates nodal points originating from crystalline symmetries\cite{mirror}. BCWO crystal structure possesses M$_x$(k$_x$=0), M$_y$(k$_y$=0), M$_z$(k$_z$=0) and three C$_4$ rotation axes, k$_x$, k$_y$ and k$_z$. Similar symmetry has been reported for Heusler alloy nodal semi-metals\cite{chatterjee, mirror}. Magnetization along (001) direction preserves only M$_z$(k$_z$=0) mirror symmetry and C$_{4z}$ rotational symmetry, and consequently, only a nodal line on the k$_z$ = 0 plane survives. We show the surviving nodal line in the k$_x$-k$_y$ in  Fig.\ref{fig4}(c) by a thin dashed line. We show only the projection in Fig.\ref{fig4}(c)  in the k$_x$-k$_y$ plane; the nodal line forms a closed loop in the E-k$_x$-k$_y$ space. 

The mirror Chern number of the system is derived by adding the contributions from each nodal line. In the presence of SOC, the mirror Chern numbers are determined using the winding numbers of Wannier centers on the mirror invariant planes. The winding number on the (K$_z$=0) mirror plane is 1, whereas on the (k$_x$=0, k$_y$=0) mirror plane it is 0. For finite SOC, the spin-orbital mixing leads to gapping of the remaining nodal line and consequently opens up a gap at these degenerate points as shown in Fig.\ref{fig4}(b) along the high-symmetry point \textit{$\Gamma$}-\textit{K}. In the Supplemental Material\cite{suppmat}, Sec. IB, we show that the gapped nodal line causes a band inversion at the nodal points close to the Fermi energy and indicates the non-trivial topological nature of the electronic band structure. The non-trivial topology of the electronic bands gives rise to non-zero Berry curvature distribution around the $E_F$. From the z-component of Berry curvature ($\Omega_z$) along the same high-symmetry path shown in Fig.\ref{fig4}(d), we find sharp peaks along the \textit{K-$\Gamma$} and L-\textit{$\Gamma$} directions and negligible contributions in the other directions. The origin of these large contributions is the small gaps in the neighborhood of the nodal line. We present the full  $\Omega_z$ distribution in the k$_x$=0 plane in\textit{ Supplemental Material}\cite{suppmat} \textit{Sec IB}. The Berry curvature induces AHC as a function of filling, as shown in the inset in Fig.\ref{fig4}(d). At $E_F$ for BCWO, we predict $\sigma_{xy}$, to be  100 S/cm.  


\begin{figure}[t]
\centering
\includegraphics[width=0.95\linewidth]{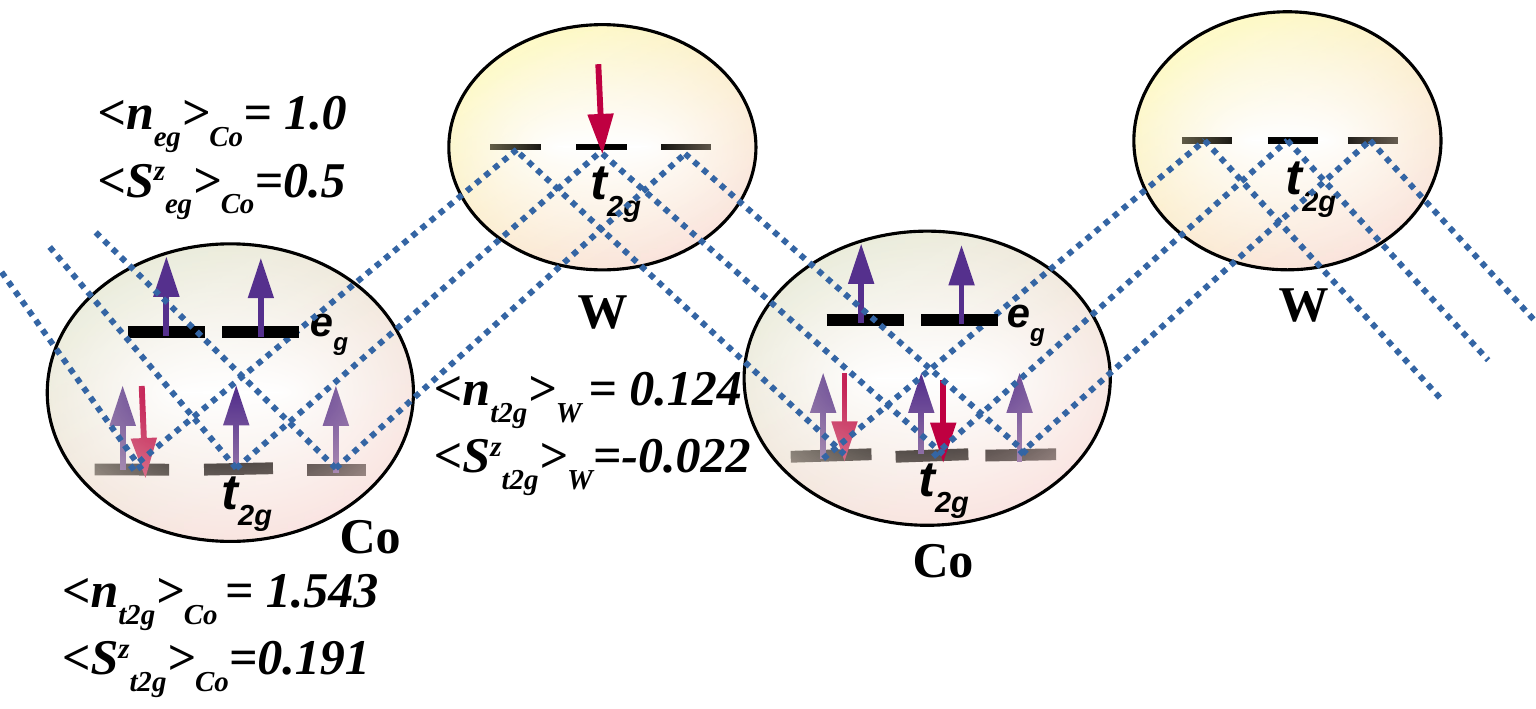}
\caption{\textbf{Exact diagonalization:}  The schematic of the ground state wavefunction and the corresponding average spin and charge densities for individual Co $t_{2g}$, Co $e_{g}$ and W $t_{2g}$ orbitals. The calculation is performed for a linear chain of two unit cells containing one Co and one W atom. The parameters are mentioned in the text.}
\label{fig03}
\end{figure}

\textit{Multi-orbital Hubbard model for BCWO:--- }  
We now consider a material-realistic model with the full $d$-manifold on Co and W retaining multi-orbital interactions with Kanamori parameterization. The onsite intra-orbital repulsion between up and down electrons $U$, inter-orbital repulsion (spin-independent) $U^\prime$, Hund's coupling preferring spin alignment ($J_H$) and pair-hopping term ($J^\prime)$ are related by $U^\prime=U-2J_H$ and $J^\prime=J_H$ \cite{kanamori}. 
To reduce computational complexity, we employ a down-folded hopping matrix $t_{\alpha,\beta}$ between Co and W orbitals, integrating out the oxygen. We further introduce SOC with coupling strength $\lambda$ on W. The full Hamiltonian, the down-folded hopping matrix, and the onsite SOC Hamiltonian are provided in \textit{Supplemental Material}\cite{suppmat} \textit{Sec II} and \textit{Sec III} respectively. We find from the down-folding that \textit{the non-zero hopping elements are only between the Co and W $t_{2g}$ orbitals.} We model the crystal field splitting between $t_{2g}$ and $e_g$ on Co and W to be $\Delta_{CF}^{Co}$=0.8eV and $\Delta_{CF}^{W}=4.0$eV respectively and set the onsite energy difference between Co and W $t_{2g}$ orbitals $\Delta_{CT}=2.5$eV. We choose $U$ in the range of 1.25eV to 2.5eV, $J_H=U/4$, and $U^\prime=U-2J_H$ for Co and vary the spin-orbit coupling (SOC) parameter $\lambda$ between 50meV to 80meV. The results below are robust for small non-zero $U$ on W and small SOC on Co. We have used the largest hopping matrix element ($t=0.2eV$) of the down-folded hopping matrix to convert the Hamiltonian parameters to energy dimensions. Our numerical conclusions discussed below also hold for $\Delta_{CF}^{Co}\in [0.5,1.0]$eV, $\Delta_{CF}^{W}\geq 3.0$eV and $\Delta_{CT}\in[2.0,3.5]$eV. The full parameter-dependent phase diagram will be reported elsewhere. 

\begin{figure}[t]
\centering
\includegraphics[width=1.0\linewidth]{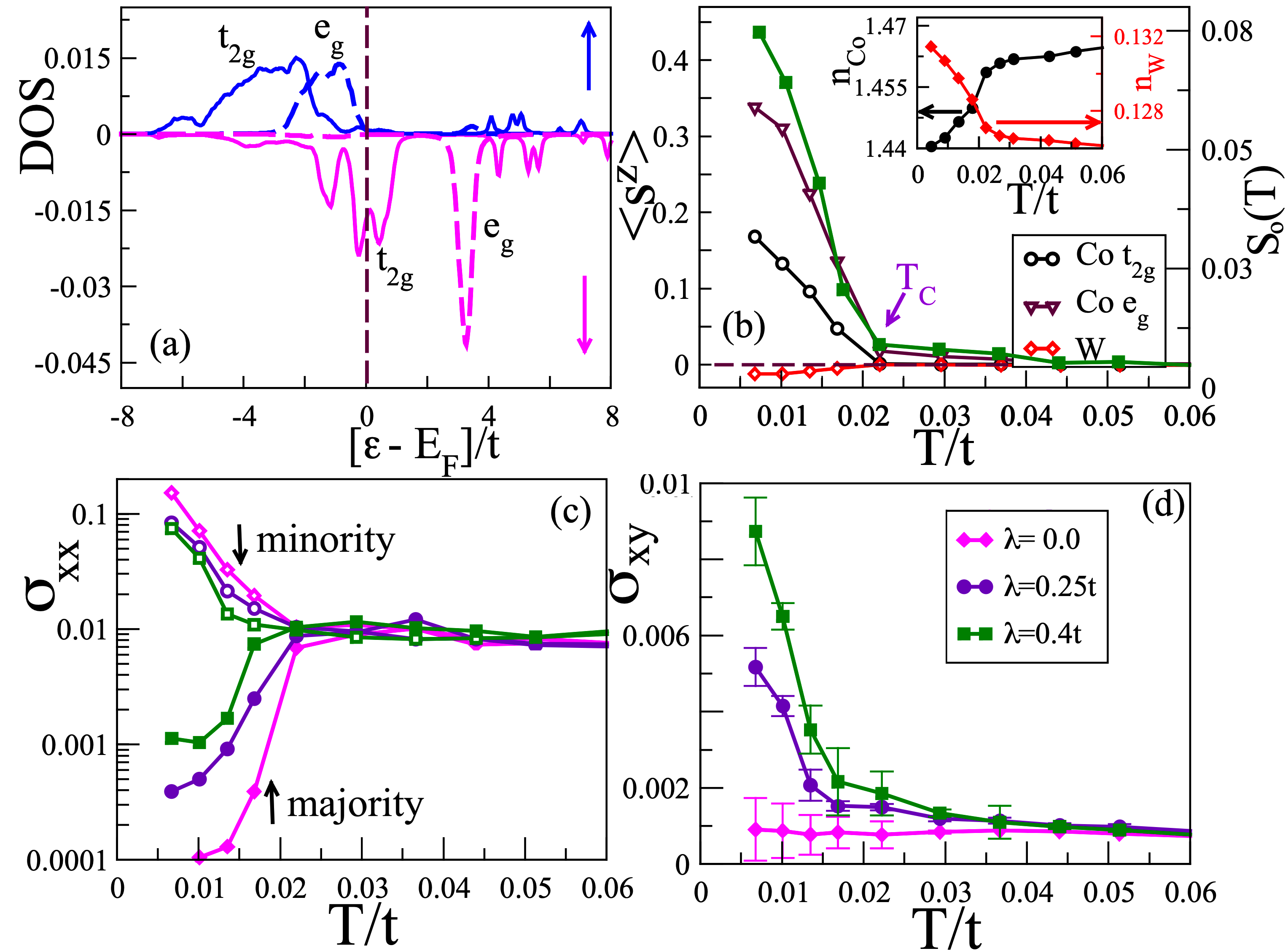}
\caption{\textbf{Finite temperature results:} (a) The orbital and spin projected DOS of Co $d$ states obtained from the 5-orbital Hubbard model study. The minority spin carriers participate in conduction and majority spin carriers are entirely gapped. (b) The system averaged expectation value of the z-component of total spin operator for Co $t_{2g}$ and $e_g$ states and for W  $t_{2g}$ states. The orbital legends are shown in (a). The green line with diamonds show the  S(0). The average charge density at Co decreases steeply at $T_C$ [see inset], where the charge density at W increases.  (c) $\sigma_{xx}(T)$  and (d) $\sigma_{xy}(T)$ for different SOC ($\lambda$)  strengths. The conductivities are measured in units of $e^2/\hbar a$ where $a=8.21$ \r{A}, is the lattice parameter. }
\label{fig2}
\end{figure}

\textit{i. Exact diagonalization:--- } 
We first perform exact diagonalization (ED) on a linear chain containing two Co-W clusters. We consider five orbitals on Co and three $t_{2g}$ orbitals on W. Given the high energy location, the $e_g$ on W is neglected, as seen from the first-principles study. For hopping, we employ the down-folded Co-W hopping elements mentioned above. We provide the details of the ED calculations in \textit{Supplemental Material}\cite{suppmat} \textit{Sec. IV}. In  Fig.\ref{fig03}  we indicate the charge density $<n_{t_{2g}}>$ and $< S^z_{t_{2g}}>$ of \textit{individual} $t_{2g}$  and  $e_g$, orbitals of Co and W. Thus, $<  S^z_{Co}>$ is 0.191$\times 3$+ 0.5$\times2$ amounting to about 1.573 and $< S^z_{W}>$=$-0.022\times 3=-0.066$. The average magnetization per unit cell is close to 3/2, as found in the first-principles calculations. In addition, we notice that the Co-W hybridization induces a finite occupation on W, unlike the nominal W ($d^0$) count. The ED results, albeit with severe size limitations, thus indicate a Hund's coupling induced high-spin Co and $t_{2g}$ minority ($\downarrow$-spin) Co-W hybridization, which leads to the half-metallic ground state. We construct a schematic representation of the ground state spin configuration in Fig.\ref{fig03}. The ED calculation, albeit with strong finite size effects, confirms the first-principles ground state.

\textit{ii. Finite temperature properties:--- } We now employ a semi-classical Monte-Carlo (s-MC) approach that provides finite-temperature properties of the multi-orbital Hubbard model\textit{ in three dimensions on 6$\times$6$\times$6 clusters with periodic boundary conditions}. The s-MC results have been demonstrated to capture physics beyond finite-temperature mean-field theory \cite{AM2,AM3}. 
We provide the methodological details in \textit{Supplemental Material}\cite{suppmat} \textit{Sec.V}. 
The low-temperature s-MC orbital and spin-resolved DOS in Fig.\ref{fig2}(a) reproduces the half-metallic ground state and agrees qualitatively with the first-principles DOS in Fig.\ref{fig1}(c). Fig.\ref{fig2}(b) shows the ($\lambda=0$) temperature evolution of the orbital resolved magnetization $\langle S^z\rangle$(T) on the left axis, and the $\mathbf{q}=0$ static magnetic structure factor ($S_0(T)$) that quantifies the ferrimagnetic order, on the right axis. At low temperature, the  $\langle S^z\rangle (T)$, for the Co $t_{2g}$  and $e_g$ orbitals are 0.54 and 0.7 respectively, while the  $\langle S^z\rangle (T)$ for W is $\sim -0.012$. The ferrimagnetic $T_c$ coincides with the concomitant vanishing of $\langle S^z\rangle$ for all orbitals. The inset in (b) shows that increasing thermal energy overcomes the energy barrier between Co and W, allowing greater occupation of the W site at the expense of the Co occupation. This thermal fluctuation-driven evolution of Co and W filling reduces the Co magnetization and suppresses the half-metallic order. 
The  ($\lambda=0$) longitudinal ($\sigma_{xx}$) and the band-topology-induced transverse  ($\sigma_{xy}$) conductivity with temperature in Fig.\ref{fig2} (c) and (d), respectively, has a clear correlation with the magnetic order shown in (b). The conductivity calculations within the Kubo-Greenwood formalism are provided in \textit{Supplemental Material} \cite{suppmat} \textit{Sec. VI}. In (c), the minority (majority) spin components of $\sigma_{xx}$ show metallic, or $d\sigma_{xx}/dT<0$ (insulating or $d\sigma_{xx}/dT>0$) behavior. The spin-dependent conductivities coincide at $T_C$. The half-metallic behavior survives for finite $\lambda$. The direction of the ferrimagnetic order allows the definition of a global z-axis of defining $\uparrow/\downarrow$, even in the presence of spin-orbital mixing SOC.
We see from Fig.\ref{fig2} (c) that the temperature where spin-polarized conductivities coincide, signaling loss of half-metallicity, is suppressed systematically with increasing $\lambda$. 
The $\sigma_{xy}$ or AHC in Fig.\ref{fig2} (d)  is zero for $\lambda=0$ within numerical resolution. For $\lambda\neq0$, we find that $\sigma_{xy}$ is finite and acquires the largest value at low temperatures. It monotonically decreases with temperature increase, with the overall magnitude being larger for greater $\lambda$. The temperature value of the inflection point in $\sigma_{xy}$ increases with decreasing $\lambda$. Compared with the $\sigma_{xx}$, there is a possibility for a temperature regime close to $T_c$ where the half-metallicity is lost, but AHC survives. Nonetheless, the $\sigma_{xy}$ is finite. However, due to the size limitations of the lattice sizes simulated, we leave the exploration of this interesting possibility for the future.
\begin{figure}[t]
\centering
\includegraphics[width=1.0\linewidth]{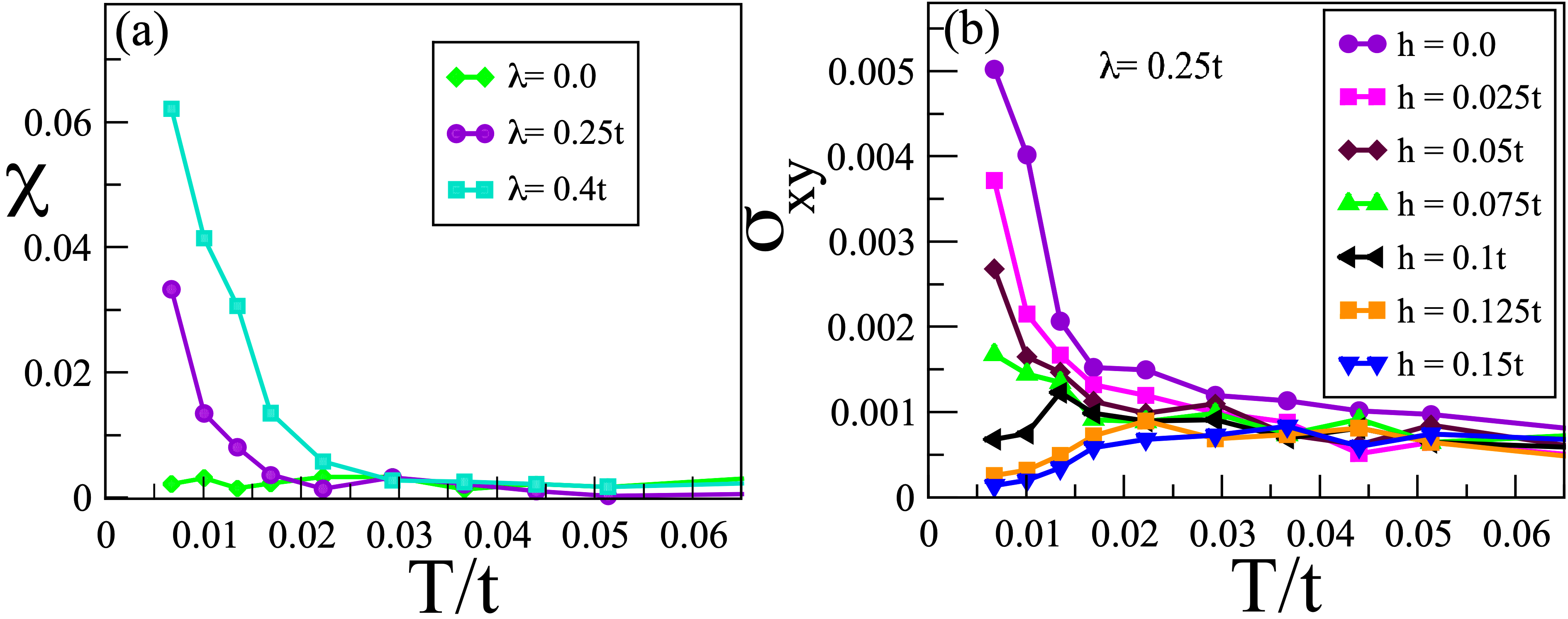}
\caption{\textbf{Spin-chirality \& transverse magnetoresponse:} We show the scalar spin chirality ($\chi$) in panel (a). The temperature dependence of $\sigma_{xy}$ in presence of external magnetic field is shown in (b).}
\label{fig5}
\end{figure}

\textit{Emergent chiral spin-texture:--- } We now demonstrate that the SOC on W induces a non-coplanarity of the Co moments. We compute the  scalar spin-chirality $\chi\equiv 1/N \sum_{i \langle j,k\rangle}\langle S_i.(S_j\times S_k)\rangle$ in the x-y plane using standard definition\cite{tex-nagaosa,tex-ahc-sq}. The spins in the above definition belong to the Co sites, and the angular brackets denote quantum and thermal averaging. $\chi
$ is calculated in individual x-y planes and summed over the planes stacked along z. 
Fig~\ref{fig5} (a) shows the temperature dependence of $\chi(T)$ and its dependence on SOC. The scalar chirality is zero, as is $\sigma_{xy}$ for $\lambda=0$, although the small contribution from intrinsic magnetization cannot be ruled out. Similarly the SOC and temperature dependence of $\chi(T)$ follow that of  $\sigma_{xy}(T)$ seen in Fig.~\ref{fig2} (d). Since scalar chirality is a measure of non-coplanarity, an external magnetic field could polarize the spins and suppress the non-coplanarity. The loss of non-coplanarity destabilizes the spin texture, which should be detrimental to $\sigma_{xy}$ if $\chi$ is indeed topological in origin. In Fig.~\ref{fig5} (b), we show a remarkable suppression of AHC with a magnetic field. The sign of $d\sigma_{xy}/dT$  at low temperatures is reversed by the magnetic field value beyond 0.075$t$. Thus the $\chi(T)$ and the $\sigma_{xy}(T)$ show a strong interdependence in accordance with literature  \cite{tex-ahc-1,tex-ahc-2}.
In contrast, the magnetic field slightly enhances the longitudinal conductivity in \textit{Supplemental Material} \cite{suppmat} \textit{Sec. VII}.

In terms of $t\sim0.2$eV, $T_c\sim 0.02t$  corresponds to 40K, within the ballpark of experimental observation ($T_C\sim 20K)$ \cite{bcwo-tc-exp} and $h_c\sim 0.1t$ for switching of $\sigma_{xy}$ for  $\lambda=50$meV  is about 95T. 
We note that the s-MC calculations are performed on small clusters due to computational complexity. As a result, they suffer from finite-size effects. So, our finite critical temperature and magnetic field scales overestimate the actual scales of the problem. Nonetheless, our results demonstrate the feasibility of such phenomena qualitatively.

\textit{Conclusion:---} In summary, we have demonstrated the fate of a nodal line semi-metal in a strongly correlated system at zero and finite temperature for the first time. Our study of BCWO reveals a half-metallic ground state with Co high-spin ferromagnetic background facilitating minority carrier delocalization in topologically non-trivial bands. SOC at W sites gaps nodal points, generating an anomalous Hall response. Using a five-orbital Hubbard model and semi-classical Monte Carlo simulations for the first time, we uncover an emergent non-coplanar spin texture at Co sites, resulting in non-zero Berry curvature and colossal transverse magneto-response controllable by an external magnetic field. This highlights the tunability of BCWO's electronic properties, which are vital for spintronic applications. Our theoretical prediction holds for other 3d-5d semi-metallic metallic DP.
Usually, spin-fermion-based half-metals have a colinear ferromagnetic or anti-ferromagnetic spin background. Here, we have modeled DP with a five-orbital Hubbard model and SOC at finite temperature for the first time. We have revealed emergent crystalline symmetry-induced scalar chirality, taking our results beyond any previous theoretical study of DP limited to spin-fermion models with classical spins coupled to fermions where transverse conductivity requires including  Dzyaloshinskii-Moriya interaction. 
 
\section*{ACKNOWLEDGMENT}
 M.K and S.C. thanks DST for funding through grant no. CRG/2020/000754. J.S thanks University Grant Commission (UGC) for Ph.D. fellowship. S.S thanks DST-INSPIRE for financial support.
\bibliography{references}

\begin{thebibliography}{36}%
\makeatletter
\providecommand \@ifxundefined [1]{%
 \@ifx{#1\undefined}
}%
\providecommand \@ifnum [1]{%
 \ifnum #1\expandafter \@firstoftwo
 \else \expandafter \@secondoftwo
 \fi
}%
\providecommand \@ifx [1]{%
 \ifx #1\expandafter \@firstoftwo
 \else \expandafter \@secondoftwo
 \fi
}%
\providecommand \natexlab [1]{#1}%
\providecommand \enquote  [1]{``#1''}%
\providecommand \bibnamefont  [1]{#1}%
\providecommand \bibfnamefont [1]{#1}%
\providecommand \citenamefont [1]{#1}%
\providecommand \href@noop [0]{\@secondoftwo}%
\providecommand \href [0]{\begingroup \@sanitize@url \@href}%
\providecommand \@href[1]{\@@startlink{#1}\@@href}%
\providecommand \@@href[1]{\endgroup#1\@@endlink}%
\providecommand \@sanitize@url [0]{\catcode `\\12\catcode `\$12\catcode `\&12\catcode `\#12\catcode `\^12\catcode `\_12\catcode `\%12\relax}%
\providecommand \@@startlink[1]{}%
\providecommand \@@endlink[0]{}%
\providecommand \url  [0]{\begingroup\@sanitize@url \@url }%
\providecommand \@url [1]{\endgroup\@href {#1}{\urlprefix }}%
\providecommand \urlprefix  [0]{URL }%
\providecommand \Eprint [0]{\href }%
\providecommand \doibase [0]{https://doi.org/}%
\providecommand \selectlanguage [0]{\@gobble}%
\providecommand \bibinfo  [0]{\@secondoftwo}%
\providecommand \bibfield  [0]{\@secondoftwo}%
\providecommand \translation [1]{[#1]}%
\providecommand \BibitemOpen [0]{}%
\providecommand \bibitemStop [0]{}%
\providecommand \bibitemNoStop [0]{.\EOS\space}%
\providecommand \EOS [0]{\spacefactor3000\relax}%
\providecommand \BibitemShut  [1]{\csname bibitem#1\endcsname}%
\let\auto@bib@innerbib\@empty
\bibitem [{\citenamefont {Fu}(2011)}]{cti-fu}%
  \BibitemOpen
  \bibfield  {author} {\bibinfo {author} {\bibfnamefont {L.}~\bibnamefont {Fu}},\ }\bibfield  {title} {\bibinfo {title} {Topological crystalline insulators},\ }\href {https://doi.org/10.1103/PhysRevLett.106.106802} {\bibfield  {journal} {\bibinfo  {journal} {Phys. Rev. Lett.}\ }\textbf {\bibinfo {volume} {106}},\ \bibinfo {pages} {106802} (\bibinfo {year} {2011})}\BibitemShut {NoStop}%
\bibitem [{\citenamefont {Burkov}(2016)}]{topo-metal-review-th}%
  \BibitemOpen
  \bibfield  {author} {\bibinfo {author} {\bibfnamefont {A.~A.}\ \bibnamefont {Burkov}},\ }\bibfield  {title} {\bibinfo {title} {Topological semimetals},\ }\href {https://doi.org/10.1038/nmat4788} {\bibfield  {journal} {\bibinfo  {journal} {Nature Materials}\ }\textbf {\bibinfo {volume} {15}},\ \bibinfo {pages} {1145} (\bibinfo {year} {2016})}\BibitemShut {NoStop}%
\bibitem [{\citenamefont {Lv}\ \emph {et~al.}(2021)\citenamefont {Lv}, \citenamefont {Qian},\ and\ \citenamefont {Ding}}]{topo-metal-review-exp}%
  \BibitemOpen
  \bibfield  {author} {\bibinfo {author} {\bibfnamefont {B.~Q.}\ \bibnamefont {Lv}}, \bibinfo {author} {\bibfnamefont {T.}~\bibnamefont {Qian}},\ and\ \bibinfo {author} {\bibfnamefont {H.}~\bibnamefont {Ding}},\ }\bibfield  {title} {\bibinfo {title} {Experimental perspective on three-dimensional topological semimetals},\ }\href {https://doi.org/10.1103/RevModPhys.93.025002} {\bibfield  {journal} {\bibinfo  {journal} {Rev. Mod. Phys.}\ }\textbf {\bibinfo {volume} {93}},\ \bibinfo {pages} {025002} (\bibinfo {year} {2021})}\BibitemShut {NoStop}%
\bibitem [{\citenamefont {Young}\ \emph {et~al.}(2012)\citenamefont {Young}, \citenamefont {Zaheer}, \citenamefont {Teo}, \citenamefont {Kane}, \citenamefont {Mele},\ and\ \citenamefont {Rappe}}]{dirac-th-1}%
  \BibitemOpen
  \bibfield  {author} {\bibinfo {author} {\bibfnamefont {S.~M.}\ \bibnamefont {Young}}, \bibinfo {author} {\bibfnamefont {S.}~\bibnamefont {Zaheer}}, \bibinfo {author} {\bibfnamefont {J.~C.~Y.}\ \bibnamefont {Teo}}, \bibinfo {author} {\bibfnamefont {C.~L.}\ \bibnamefont {Kane}}, \bibinfo {author} {\bibfnamefont {E.~J.}\ \bibnamefont {Mele}},\ and\ \bibinfo {author} {\bibfnamefont {A.~M.}\ \bibnamefont {Rappe}},\ }\bibfield  {title} {\bibinfo {title} {Dirac semimetal in three dimensions},\ }\href {https://doi.org/10.1103/PhysRevLett.108.140405} {\bibfield  {journal} {\bibinfo  {journal} {Phys. Rev. Lett.}\ }\textbf {\bibinfo {volume} {108}},\ \bibinfo {pages} {140405} (\bibinfo {year} {2012})}\BibitemShut {NoStop}%
\bibitem [{\citenamefont {Wang}\ \emph {et~al.}(2012)\citenamefont {Wang}, \citenamefont {Sun}, \citenamefont {Chen}, \citenamefont {Franchini}, \citenamefont {Xu}, \citenamefont {Weng}, \citenamefont {Dai},\ and\ \citenamefont {Fang}}]{dirac-th-2}%
  \BibitemOpen
  \bibfield  {author} {\bibinfo {author} {\bibfnamefont {Z.}~\bibnamefont {Wang}}, \bibinfo {author} {\bibfnamefont {Y.}~\bibnamefont {Sun}}, \bibinfo {author} {\bibfnamefont {X.-Q.}\ \bibnamefont {Chen}}, \bibinfo {author} {\bibfnamefont {C.}~\bibnamefont {Franchini}}, \bibinfo {author} {\bibfnamefont {G.}~\bibnamefont {Xu}}, \bibinfo {author} {\bibfnamefont {H.}~\bibnamefont {Weng}}, \bibinfo {author} {\bibfnamefont {X.}~\bibnamefont {Dai}},\ and\ \bibinfo {author} {\bibfnamefont {Z.}~\bibnamefont {Fang}},\ }\bibfield  {title} {\bibinfo {title} {Dirac semimetal and topological phase transitions in ${A}_{3}${B}i ({A}={Na}, {K}, {Rb})},\ }\href {https://doi.org/10.1103/PhysRevB.85.195320} {\bibfield  {journal} {\bibinfo  {journal} {Phys. Rev. B}\ }\textbf {\bibinfo {volume} {85}},\ \bibinfo {pages} {195320} (\bibinfo {year} {2012})}\BibitemShut {NoStop}%
\bibitem [{\citenamefont {Wang}\ \emph {et~al.}(2013)\citenamefont {Wang}, \citenamefont {Weng}, \citenamefont {Wu}, \citenamefont {Dai},\ and\ \citenamefont {Fang}}]{dirac-th-3}%
  \BibitemOpen
  \bibfield  {author} {\bibinfo {author} {\bibfnamefont {Z.}~\bibnamefont {Wang}}, \bibinfo {author} {\bibfnamefont {H.}~\bibnamefont {Weng}}, \bibinfo {author} {\bibfnamefont {Q.}~\bibnamefont {Wu}}, \bibinfo {author} {\bibfnamefont {X.}~\bibnamefont {Dai}},\ and\ \bibinfo {author} {\bibfnamefont {Z.}~\bibnamefont {Fang}},\ }\bibfield  {title} {\bibinfo {title} {Three-dimensional dirac semimetal and quantum transport in ${Cd}_{3}{As}_{2}$},\ }\href {https://doi.org/10.1103/PhysRevB.88.125427} {\bibfield  {journal} {\bibinfo  {journal} {Phys. Rev. B}\ }\textbf {\bibinfo {volume} {88}},\ \bibinfo {pages} {125427} (\bibinfo {year} {2013})}\BibitemShut {NoStop}%
\bibitem [{\citenamefont {Li}\ \emph {et~al.}(2020)\citenamefont {Li}, \citenamefont {Zhang}, \citenamefont {Wang}, \citenamefont {Huang}, \citenamefont {Gu},\ and\ \citenamefont {Duan}}]{semi-metal-dft-review-2}%
  \BibitemOpen
  \bibfield  {author} {\bibinfo {author} {\bibfnamefont {J.}~\bibnamefont {Li}}, \bibinfo {author} {\bibfnamefont {Z.}~\bibnamefont {Zhang}}, \bibinfo {author} {\bibfnamefont {C.}~\bibnamefont {Wang}}, \bibinfo {author} {\bibfnamefont {H.}~\bibnamefont {Huang}}, \bibinfo {author} {\bibfnamefont {B.-L.}\ \bibnamefont {Gu}},\ and\ \bibinfo {author} {\bibfnamefont {W.}~\bibnamefont {Duan}},\ }\bibfield  {title} {\bibinfo {title} {{Topological semimetals from the perspective of first-principles calculations}},\ }\href {https://doi.org/10.1063/5.0025396} {\bibfield  {journal} {\bibinfo  {journal} {Journal of Applied Physics}\ }\textbf {\bibinfo {volume} {128}},\ \bibinfo {pages} {191101} (\bibinfo {year} {2020})}\BibitemShut {NoStop}%
\bibitem [{\citenamefont {Burkov}\ and\ \citenamefont {Balents}(2011)}]{Weyl-th-1}%
  \BibitemOpen
  \bibfield  {author} {\bibinfo {author} {\bibfnamefont {A.~A.}\ \bibnamefont {Burkov}}\ and\ \bibinfo {author} {\bibfnamefont {L.}~\bibnamefont {Balents}},\ }\bibfield  {title} {\bibinfo {title} {Weyl semimetal in a topological insulator multilayer},\ }\href {https://doi.org/10.1103/PhysRevLett.107.127205} {\bibfield  {journal} {\bibinfo  {journal} {Phys. Rev. Lett.}\ }\textbf {\bibinfo {volume} {107}},\ \bibinfo {pages} {127205} (\bibinfo {year} {2011})}\BibitemShut {NoStop}%
\bibitem [{\citenamefont {Xu}\ \emph {et~al.}(2011)\citenamefont {Xu}, \citenamefont {Weng}, \citenamefont {Wang}, \citenamefont {Dai},\ and\ \citenamefont {Fang}}]{Weyl-th-2}%
  \BibitemOpen
  \bibfield  {author} {\bibinfo {author} {\bibfnamefont {G.}~\bibnamefont {Xu}}, \bibinfo {author} {\bibfnamefont {H.}~\bibnamefont {Weng}}, \bibinfo {author} {\bibfnamefont {Z.}~\bibnamefont {Wang}}, \bibinfo {author} {\bibfnamefont {X.}~\bibnamefont {Dai}},\ and\ \bibinfo {author} {\bibfnamefont {Z.}~\bibnamefont {Fang}},\ }\bibfield  {title} {\bibinfo {title} {Chern semimetal and the quantized anomalous hall effect in ${Hg}{Cr}_{2}{Se}_{4}$},\ }\href {https://doi.org/10.1103/PhysRevLett.107.186806} {\bibfield  {journal} {\bibinfo  {journal} {Phys. Rev. Lett.}\ }\textbf {\bibinfo {volume} {107}},\ \bibinfo {pages} {186806} (\bibinfo {year} {2011})}\BibitemShut {NoStop}%
\bibitem [{\citenamefont {Liu}\ \emph {et~al.}(2014)\citenamefont {Liu}, \citenamefont {Zhou}, \citenamefont {Zhang}, \citenamefont {Wang}, \citenamefont {Weng}, \citenamefont {Prabhakaran}, \citenamefont {Mo}, \citenamefont {Shen}, \citenamefont {Fang}, \citenamefont {Dai}, \citenamefont {Hussain},\ and\ \citenamefont {Chen}}]{dirac-exp-1}%
  \BibitemOpen
  \bibfield  {author} {\bibinfo {author} {\bibfnamefont {Z.~K.}\ \bibnamefont {Liu}}, \bibinfo {author} {\bibfnamefont {B.}~\bibnamefont {Zhou}}, \bibinfo {author} {\bibfnamefont {Y.}~\bibnamefont {Zhang}}, \bibinfo {author} {\bibfnamefont {Z.~J.}\ \bibnamefont {Wang}}, \bibinfo {author} {\bibfnamefont {H.~M.}\ \bibnamefont {Weng}}, \bibinfo {author} {\bibfnamefont {D.}~\bibnamefont {Prabhakaran}}, \bibinfo {author} {\bibfnamefont {S.-K.}\ \bibnamefont {Mo}}, \bibinfo {author} {\bibfnamefont {Z.~X.}\ \bibnamefont {Shen}}, \bibinfo {author} {\bibfnamefont {Z.}~\bibnamefont {Fang}}, \bibinfo {author} {\bibfnamefont {X.}~\bibnamefont {Dai}}, \bibinfo {author} {\bibfnamefont {Z.}~\bibnamefont {Hussain}},\ and\ \bibinfo {author} {\bibfnamefont {Y.~L.}\ \bibnamefont {Chen}},\ }\bibfield  {title} {\bibinfo {title} {Discovery of a three-dimensional topological dirac semimetal, ${Na}_{3}{Bi}$},\ }\href {https://doi.org/10.1126/science.1245085} {\bibfield  {journal} {\bibinfo  {journal} {Science}\ }\textbf {\bibinfo
  {volume} {343}},\ \bibinfo {pages} {864} (\bibinfo {year} {2014})}\BibitemShut {NoStop}%
\bibitem [{\citenamefont {Xu}\ \emph {et~al.}(2015)\citenamefont {Xu}, \citenamefont {Belopolski}, \citenamefont {Alidoust}, \citenamefont {Neupane}, \citenamefont {Bian}, \citenamefont {Zhang}, \citenamefont {Sankar}, \citenamefont {Chang}, \citenamefont {Yuan}, \citenamefont {Lee}, \citenamefont {Huang}, \citenamefont {Zheng}, \citenamefont {Ma}, \citenamefont {Sanchez}, \citenamefont {Wang}, \citenamefont {Bansil}, \citenamefont {Chou}, \citenamefont {Shibayev}, \citenamefont {Lin}, \citenamefont {Jia},\ and\ \citenamefont {Hasan}}]{Weyl-exp-1}%
  \BibitemOpen
  \bibfield  {author} {\bibinfo {author} {\bibfnamefont {S.-Y.}\ \bibnamefont {Xu}}, \bibinfo {author} {\bibfnamefont {I.}~\bibnamefont {Belopolski}}, \bibinfo {author} {\bibfnamefont {N.}~\bibnamefont {Alidoust}}, \bibinfo {author} {\bibfnamefont {M.}~\bibnamefont {Neupane}}, \bibinfo {author} {\bibfnamefont {G.}~\bibnamefont {Bian}}, \bibinfo {author} {\bibfnamefont {C.}~\bibnamefont {Zhang}}, \bibinfo {author} {\bibfnamefont {R.}~\bibnamefont {Sankar}}, \bibinfo {author} {\bibfnamefont {G.}~\bibnamefont {Chang}}, \bibinfo {author} {\bibfnamefont {Z.}~\bibnamefont {Yuan}}, \bibinfo {author} {\bibfnamefont {C.-C.}\ \bibnamefont {Lee}}, \bibinfo {author} {\bibfnamefont {S.-M.}\ \bibnamefont {Huang}}, \bibinfo {author} {\bibfnamefont {H.}~\bibnamefont {Zheng}}, \bibinfo {author} {\bibfnamefont {J.}~\bibnamefont {Ma}}, \bibinfo {author} {\bibfnamefont {D.~S.}\ \bibnamefont {Sanchez}}, \bibinfo {author} {\bibfnamefont {B.}~\bibnamefont {Wang}}, \bibinfo {author} {\bibfnamefont {A.}~\bibnamefont {Bansil}},
  \bibinfo {author} {\bibfnamefont {F.}~\bibnamefont {Chou}}, \bibinfo {author} {\bibfnamefont {P.~P.}\ \bibnamefont {Shibayev}}, \bibinfo {author} {\bibfnamefont {H.}~\bibnamefont {Lin}}, \bibinfo {author} {\bibfnamefont {S.}~\bibnamefont {Jia}},\ and\ \bibinfo {author} {\bibfnamefont {M.~Z.}\ \bibnamefont {Hasan}},\ }\bibfield  {title} {\bibinfo {title} {Discovery of a weyl fermion semimetal and topological fermi arcs},\ }\href {https://doi.org/10.1126/science.aaa9297} {\bibfield  {journal} {\bibinfo  {journal} {Science}\ }\textbf {\bibinfo {volume} {349}},\ \bibinfo {pages} {613} (\bibinfo {year} {2015})}\BibitemShut {NoStop}%
\bibitem [{\citenamefont {Lv}\ \emph {et~al.}(2015)\citenamefont {Lv}, \citenamefont {Weng}, \citenamefont {Fu}, \citenamefont {Wang}, \citenamefont {Miao}, \citenamefont {Ma}, \citenamefont {Richard}, \citenamefont {Huang}, \citenamefont {Zhao}, \citenamefont {Chen}, \citenamefont {Fang}, \citenamefont {Dai}, \citenamefont {Qian},\ and\ \citenamefont {Ding}}]{Weyl-exp-2}%
  \BibitemOpen
  \bibfield  {author} {\bibinfo {author} {\bibfnamefont {B.~Q.}\ \bibnamefont {Lv}}, \bibinfo {author} {\bibfnamefont {H.~M.}\ \bibnamefont {Weng}}, \bibinfo {author} {\bibfnamefont {B.~B.}\ \bibnamefont {Fu}}, \bibinfo {author} {\bibfnamefont {X.~P.}\ \bibnamefont {Wang}}, \bibinfo {author} {\bibfnamefont {H.}~\bibnamefont {Miao}}, \bibinfo {author} {\bibfnamefont {J.}~\bibnamefont {Ma}}, \bibinfo {author} {\bibfnamefont {P.}~\bibnamefont {Richard}}, \bibinfo {author} {\bibfnamefont {X.~C.}\ \bibnamefont {Huang}}, \bibinfo {author} {\bibfnamefont {L.~X.}\ \bibnamefont {Zhao}}, \bibinfo {author} {\bibfnamefont {G.~F.}\ \bibnamefont {Chen}}, \bibinfo {author} {\bibfnamefont {Z.}~\bibnamefont {Fang}}, \bibinfo {author} {\bibfnamefont {X.}~\bibnamefont {Dai}}, \bibinfo {author} {\bibfnamefont {T.}~\bibnamefont {Qian}},\ and\ \bibinfo {author} {\bibfnamefont {H.}~\bibnamefont {Ding}},\ }\bibfield  {title} {\bibinfo {title} {Experimental discovery of weyl semimetal {Ta}{As}},\ }\href
  {https://doi.org/10.1103/PhysRevX.5.031013} {\bibfield  {journal} {\bibinfo  {journal} {Phys. Rev. X}\ }\textbf {\bibinfo {volume} {5}},\ \bibinfo {pages} {031013} (\bibinfo {year} {2015})}\BibitemShut {NoStop}%
\bibitem [{\citenamefont {Hsieh}\ \emph {et~al.}(2012)\citenamefont {Hsieh}, \citenamefont {Lin}, \citenamefont {Liu}, \citenamefont {Duan}, \citenamefont {Bansil},\ and\ \citenamefont {Fu}}]{mirror-chern-1}%
  \BibitemOpen
  \bibfield  {author} {\bibinfo {author} {\bibfnamefont {T.~H.}\ \bibnamefont {Hsieh}}, \bibinfo {author} {\bibfnamefont {H.}~\bibnamefont {Lin}}, \bibinfo {author} {\bibfnamefont {J.}~\bibnamefont {Liu}}, \bibinfo {author} {\bibfnamefont {W.}~\bibnamefont {Duan}}, \bibinfo {author} {\bibfnamefont {A.}~\bibnamefont {Bansil}},\ and\ \bibinfo {author} {\bibfnamefont {L.}~\bibnamefont {Fu}},\ }\bibfield  {title} {\bibinfo {title} {Topological crystalline insulators in the {SnTe} material class},\ }\href@noop {} {\bibfield  {journal} {\bibinfo  {journal} {Nature Communications}\ }\textbf {\bibinfo {volume} {3}},\ \bibinfo {pages} {982} (\bibinfo {year} {2012})}\BibitemShut {NoStop}%
\bibitem [{\citenamefont {Wang}\ \emph {et~al.}(2024)\citenamefont {Wang}, \citenamefont {Cui}, \citenamefont {Zhang}, \citenamefont {Wang}, \citenamefont {Yu}, \citenamefont {Liu},\ and\ \citenamefont {Yao}}]{mirror-chern-2}%
  \BibitemOpen
  \bibfield  {author} {\bibinfo {author} {\bibfnamefont {Y.}~\bibnamefont {Wang}}, \bibinfo {author} {\bibfnamefont {C.}~\bibnamefont {Cui}}, \bibinfo {author} {\bibfnamefont {R.-W.}\ \bibnamefont {Zhang}}, \bibinfo {author} {\bibfnamefont {X.}~\bibnamefont {Wang}}, \bibinfo {author} {\bibfnamefont {Z.-M.}\ \bibnamefont {Yu}}, \bibinfo {author} {\bibfnamefont {G.-B.}\ \bibnamefont {Liu}},\ and\ \bibinfo {author} {\bibfnamefont {Y.}~\bibnamefont {Yao}},\ }\bibfield  {title} {\bibinfo {title} {Mirror real chern insulator in two and three dimensions},\ }\href {https://doi.org/10.1103/PhysRevB.109.195101} {\bibfield  {journal} {\bibinfo  {journal} {Phys. Rev. B}\ }\textbf {\bibinfo {volume} {109}},\ \bibinfo {pages} {195101} (\bibinfo {year} {2024})}\BibitemShut {NoStop}%
\bibitem [{\citenamefont {Chang}\ \emph {et~al.}(2016)\citenamefont {Chang}, \citenamefont {Xu}, \citenamefont {Zheng}, \citenamefont {Singh}, \citenamefont {Hsu}, \citenamefont {Bian}, \citenamefont {Alidoust}, \citenamefont {Belopolski}, \citenamefont {Sanchez}, \citenamefont {Zhang} \emph {et~al.}}]{mirror}%
  \BibitemOpen
  \bibfield  {author} {\bibinfo {author} {\bibfnamefont {G.}~\bibnamefont {Chang}}, \bibinfo {author} {\bibfnamefont {S.-Y.}\ \bibnamefont {Xu}}, \bibinfo {author} {\bibfnamefont {H.}~\bibnamefont {Zheng}}, \bibinfo {author} {\bibfnamefont {B.}~\bibnamefont {Singh}}, \bibinfo {author} {\bibfnamefont {C.-H.}\ \bibnamefont {Hsu}}, \bibinfo {author} {\bibfnamefont {G.}~\bibnamefont {Bian}}, \bibinfo {author} {\bibfnamefont {N.}~\bibnamefont {Alidoust}}, \bibinfo {author} {\bibfnamefont {I.}~\bibnamefont {Belopolski}}, \bibinfo {author} {\bibfnamefont {D.~S.}\ \bibnamefont {Sanchez}}, \bibinfo {author} {\bibfnamefont {S.}~\bibnamefont {Zhang}}, \emph {et~al.},\ }\bibfield  {title} {\bibinfo {title} {Room-temperature magnetic topological weyl fermion and nodal line semimetal states in half-metallic heusler ${Co}_{2}{Ti}{X}$ ({X}= {Si}, {Ge}, or {Sn})},\ }\href@noop {} {\bibfield  {journal} {\bibinfo  {journal} {Scientific reports}\ }\textbf {\bibinfo {volume} {6}},\ \bibinfo {pages} {38839} (\bibinfo {year}
  {2016})}\BibitemShut {NoStop}%
\bibitem [{\citenamefont {An}\ \emph {et~al.}(2022)\citenamefont {An}, \citenamefont {Zhao}, \citenamefont {Wang},\ and\ \citenamefont {Zhang}}]{application-hall}%
  \BibitemOpen
  \bibfield  {author} {\bibinfo {author} {\bibfnamefont {P.}~\bibnamefont {An}}, \bibinfo {author} {\bibfnamefont {H.}~\bibnamefont {Zhao}}, \bibinfo {author} {\bibfnamefont {R.}~\bibnamefont {Wang}},\ and\ \bibinfo {author} {\bibfnamefont {C.}~\bibnamefont {Zhang}},\ }\bibfield  {title} {\bibinfo {title} {The recent progress and the state-of-art applications for different types of hall effect},\ }\href {https://doi.org/10.1088/1742-6596/2386/1/012061} {\bibfield  {journal} {\bibinfo  {journal} {Journal of Physics: Conference Series}\ }\textbf {\bibinfo {volume} {2386}},\ \bibinfo {pages} {012061} (\bibinfo {year} {2022})}\BibitemShut {NoStop}%
\bibitem [{\citenamefont {Yang}(2016)}]{application-spin}%
  \BibitemOpen
  \bibfield  {author} {\bibinfo {author} {\bibfnamefont {S.~A.}\ \bibnamefont {Yang}},\ }\bibfield  {title} {\bibinfo {title} {Dirac and weyl materials: Fundamental aspects and some spintronics applications},\ }\href {https://doi.org/10.1142/S2010324716400038} {\bibfield  {journal} {\bibinfo  {journal} {SPIN}\ }\textbf {\bibinfo {volume} {06}},\ \bibinfo {pages} {1640003} (\bibinfo {year} {2016})},\ \Eprint {https://arxiv.org/abs/https://doi.org/10.1142/S2010324716400038} {https://doi.org/10.1142/S2010324716400038} \BibitemShut {NoStop}%
\bibitem [{\citenamefont {Nayak}\ \emph {et~al.}(2008)\citenamefont {Nayak}, \citenamefont {Simon}, \citenamefont {Stern}, \citenamefont {Freedman},\ and\ \citenamefont {Das~Sarma}}]{application-quant}%
  \BibitemOpen
  \bibfield  {author} {\bibinfo {author} {\bibfnamefont {C.}~\bibnamefont {Nayak}}, \bibinfo {author} {\bibfnamefont {S.~H.}\ \bibnamefont {Simon}}, \bibinfo {author} {\bibfnamefont {A.}~\bibnamefont {Stern}}, \bibinfo {author} {\bibfnamefont {M.}~\bibnamefont {Freedman}},\ and\ \bibinfo {author} {\bibfnamefont {S.}~\bibnamefont {Das~Sarma}},\ }\bibfield  {title} {\bibinfo {title} {Non-abelian anyons and topological quantum computation},\ }\href {https://doi.org/10.1103/RevModPhys.80.1083} {\bibfield  {journal} {\bibinfo  {journal} {Rev. Mod. Phys.}\ }\textbf {\bibinfo {volume} {80}},\ \bibinfo {pages} {1083} (\bibinfo {year} {2008})}\BibitemShut {NoStop}%
\bibitem [{\citenamefont {Katsnelson}\ \emph {et~al.}(2008)\citenamefont {Katsnelson}, \citenamefont {Irkhin}, \citenamefont {Chioncel}, \citenamefont {Lichtenstein},\ and\ \citenamefont {de~Groot}}]{half-metal-rmp}%
  \BibitemOpen
  \bibfield  {author} {\bibinfo {author} {\bibfnamefont {M.~I.}\ \bibnamefont {Katsnelson}}, \bibinfo {author} {\bibfnamefont {V.~Y.}\ \bibnamefont {Irkhin}}, \bibinfo {author} {\bibfnamefont {L.}~\bibnamefont {Chioncel}}, \bibinfo {author} {\bibfnamefont {A.~I.}\ \bibnamefont {Lichtenstein}},\ and\ \bibinfo {author} {\bibfnamefont {R.~A.}\ \bibnamefont {de~Groot}},\ }\bibfield  {title} {\bibinfo {title} {Half-metallic ferromagnets: From band structure to many-body effects},\ }\href {https://doi.org/10.1103/RevModPhys.80.315} {\bibfield  {journal} {\bibinfo  {journal} {Rev. Mod. Phys.}\ }\textbf {\bibinfo {volume} {80}},\ \bibinfo {pages} {315} (\bibinfo {year} {2008})}\BibitemShut {NoStop}%
\bibitem [{\citenamefont {Serrate}\ \emph {et~al.}(2006)\citenamefont {Serrate}, \citenamefont {Teresa},\ and\ \citenamefont {Ibarra}}]{half-metal-review-1}%
  \BibitemOpen
  \bibfield  {author} {\bibinfo {author} {\bibfnamefont {D.}~\bibnamefont {Serrate}}, \bibinfo {author} {\bibfnamefont {J.~M.~D.}\ \bibnamefont {Teresa}},\ and\ \bibinfo {author} {\bibfnamefont {M.~R.}\ \bibnamefont {Ibarra}},\ }\bibfield  {title} {\bibinfo {title} {Double perovskites with ferromagnetism above room temperature},\ }\href {https://doi.org/10.1088/0953-8984/19/2/023201} {\bibfield  {journal} {\bibinfo  {journal} {Journal of Physics: Condensed Matter}\ }\textbf {\bibinfo {volume} {19}},\ \bibinfo {pages} {023201} (\bibinfo {year} {2006})}\BibitemShut {NoStop}%
\bibitem [{\citenamefont {Nair}\ \emph {et~al.}(2014)\citenamefont {Nair}, \citenamefont {Pradheesh}, \citenamefont {Xiao}, \citenamefont {Cherian}, \citenamefont {Elizabeth}, \citenamefont {Hansen}, \citenamefont {Chatterji},\ and\ \citenamefont {Bruckel}}]{antisite-dis-old}%
  \BibitemOpen
  \bibfield  {author} {\bibinfo {author} {\bibfnamefont {H.~S.}\ \bibnamefont {Nair}}, \bibinfo {author} {\bibfnamefont {R.}~\bibnamefont {Pradheesh}}, \bibinfo {author} {\bibfnamefont {Y.}~\bibnamefont {Xiao}}, \bibinfo {author} {\bibfnamefont {D.}~\bibnamefont {Cherian}}, \bibinfo {author} {\bibfnamefont {S.}~\bibnamefont {Elizabeth}}, \bibinfo {author} {\bibfnamefont {T.}~\bibnamefont {Hansen}}, \bibinfo {author} {\bibfnamefont {T.}~\bibnamefont {Chatterji}},\ and\ \bibinfo {author} {\bibfnamefont {T.}~\bibnamefont {Bruckel}},\ }\bibfield  {title} {\bibinfo {title} {{Magnetization-steps in ${Y}_{2}{C}o{M}n{O}_{6}$ double perovskite: The role of antisite disorder}},\ }\href {https://doi.org/10.1063/1.4896399} {\bibfield  {journal} {\bibinfo  {journal} {Journal of Applied Physics}\ }\textbf {\bibinfo {volume} {116}},\ \bibinfo {pages} {123907} (\bibinfo {year} {2014})}\BibitemShut {NoStop}%
\bibitem [{\citenamefont {Plumb}\ \emph {et~al.}(2013)\citenamefont {Plumb}, \citenamefont {Cook}, \citenamefont {Clancy}, \citenamefont {Kolesnikov}, \citenamefont {Jeon}, \citenamefont {Noh}, \citenamefont {Paramekanti},\ and\ \citenamefont {Kim}}]{antisite-dis-1}%
  \BibitemOpen
  \bibfield  {author} {\bibinfo {author} {\bibfnamefont {K.~W.}\ \bibnamefont {Plumb}}, \bibinfo {author} {\bibfnamefont {A.~M.}\ \bibnamefont {Cook}}, \bibinfo {author} {\bibfnamefont {J.~P.}\ \bibnamefont {Clancy}}, \bibinfo {author} {\bibfnamefont {A.~I.}\ \bibnamefont {Kolesnikov}}, \bibinfo {author} {\bibfnamefont {B.~C.}\ \bibnamefont {Jeon}}, \bibinfo {author} {\bibfnamefont {T.~W.}\ \bibnamefont {Noh}}, \bibinfo {author} {\bibfnamefont {A.}~\bibnamefont {Paramekanti}},\ and\ \bibinfo {author} {\bibfnamefont {Y.-J.}\ \bibnamefont {Kim}},\ }\bibfield  {title} {\bibinfo {title} {Neutron scattering study of magnetic excitations in a 5$d$-based double-perovskite {B}a${}_{2}${F}e{R}e{O}${}_{6}$},\ }\href {https://doi.org/10.1103/PhysRevB.87.184412} {\bibfield  {journal} {\bibinfo  {journal} {Phys. Rev. B}\ }\textbf {\bibinfo {volume} {87}},\ \bibinfo {pages} {184412} (\bibinfo {year} {2013})}\BibitemShut {NoStop}%
\bibitem [{\citenamefont {Gao}\ \emph {et~al.}(2019)\citenamefont {Gao}, \citenamefont {Venderbos}, \citenamefont {Kim},\ and\ \citenamefont {Rappe}}]{semi-metal-dft-review-1}%
  \BibitemOpen
  \bibfield  {author} {\bibinfo {author} {\bibfnamefont {H.}~\bibnamefont {Gao}}, \bibinfo {author} {\bibfnamefont {J.~W.}\ \bibnamefont {Venderbos}}, \bibinfo {author} {\bibfnamefont {Y.}~\bibnamefont {Kim}},\ and\ \bibinfo {author} {\bibfnamefont {A.~M.}\ \bibnamefont {Rappe}},\ }\bibfield  {title} {\bibinfo {title} {Topological semimetals from first principles},\ }\href {https://doi.org/https://doi.org/10.1146/annurev-matsci-070218-010049} {\bibfield  {journal} {\bibinfo  {journal} {Annual Review of Materials Research}\ }\textbf {\bibinfo {volume} {49}},\ \bibinfo {pages} {153} (\bibinfo {year} {2019})}\BibitemShut {NoStop}%
\bibitem [{\citenamefont {Cook}\ and\ \citenamefont {Paramekanti}(2014)}]{bfro-bfwo-prl-cook}%
  \BibitemOpen
  \bibfield  {author} {\bibinfo {author} {\bibfnamefont {A.~M.}\ \bibnamefont {Cook}}\ and\ \bibinfo {author} {\bibfnamefont {A.}~\bibnamefont {Paramekanti}},\ }\bibfield  {title} {\bibinfo {title} {Double perovskite heterostructures: Magnetism, chern bands, and chern insulators},\ }\href {https://doi.org/10.1103/PhysRevLett.113.077203} {\bibfield  {journal} {\bibinfo  {journal} {Phys. Rev. Lett.}\ }\textbf {\bibinfo {volume} {113}},\ \bibinfo {pages} {077203} (\bibinfo {year} {2014})}\BibitemShut {NoStop}%
\bibitem [{\citenamefont {Cook}\ and\ \citenamefont {Paramekanti}(2013)}]{bfr-prb-cook}%
  \BibitemOpen
  \bibfield  {author} {\bibinfo {author} {\bibfnamefont {A.}~\bibnamefont {Cook}}\ and\ \bibinfo {author} {\bibfnamefont {A.}~\bibnamefont {Paramekanti}},\ }\bibfield  {title} {\bibinfo {title} {Theory of metallic double perovskites with spin-orbit coupling and strong correlations: Application to ferrimagnetic {B}a${}_{2}${F}e{R}e{O}${}_{6}$},\ }\href {https://doi.org/10.1103/PhysRevB.88.235102} {\bibfield  {journal} {\bibinfo  {journal} {Phys. Rev. B}\ }\textbf {\bibinfo {volume} {88}},\ \bibinfo {pages} {235102} (\bibinfo {year} {2013})}\BibitemShut {NoStop}%
\bibitem [{\citenamefont {Noce}\ and\ \citenamefont {Romano}(2014)}]{kanamori}%
  \BibitemOpen
  \bibfield  {author} {\bibinfo {author} {\bibfnamefont {C.}~\bibnamefont {Noce}}\ and\ \bibinfo {author} {\bibfnamefont {A.}~\bibnamefont {Romano}},\ }\bibfield  {title} {\bibinfo {title} {{Rotationally invariant parametrization of Coulomb interactions in multi-orbital Hubbard models}},\ }\href {https://doi.org/10.1002/pssb.201350148} {\bibfield  {journal} {\bibinfo  {journal} {physica status solidi (b)}\ }\textbf {\bibinfo {volume} {251}},\ \bibinfo {pages} {907} (\bibinfo {year} {2014})}\BibitemShut {NoStop}%
\bibitem [{sup()}]{suppmat}%
  \BibitemOpen
  \href@noop {} {}\bibinfo {note} {Supplemental Material: We here provide supplemental explanations and data on the following topics in relation to the main text: I. Details of the first principles methodologies and Wannierization leading to down-folded Hamiltonian. II. Details of the down-folded Hamiltonian and the anomalous Hall conductivity within first-principles calculations. III. contains the details of the multi-orbital and spin-orbit Hamiltonian. IV discussed the exact diagonalization results. In V we discuss the semi-classical Monte-Carlo approach and the Kubo-Greenwood formalism for conductivity is discussed in VI. In VII we provide results on the field dependence of the longitudinal conductivity as a function of temperature.}\BibitemShut {Stop}%
\bibitem [{\citenamefont {Ray}\ \emph {et~al.}(2014)\citenamefont {Ray}, \citenamefont {Himanshu}, \citenamefont {Brajesh}, \citenamefont {Choudhary}, \citenamefont {Bandyopadhyay}, \citenamefont {Sen}, \citenamefont {Kumar},\ and\ \citenamefont {Sinha}}]{bwco-hm-dft-old}%
  \BibitemOpen
  \bibfield  {author} {\bibinfo {author} {\bibfnamefont {R.}~\bibnamefont {Ray}}, \bibinfo {author} {\bibfnamefont {A.~K.}\ \bibnamefont {Himanshu}}, \bibinfo {author} {\bibfnamefont {K.}~\bibnamefont {Brajesh}}, \bibinfo {author} {\bibfnamefont {B.~K.}\ \bibnamefont {Choudhary}}, \bibinfo {author} {\bibfnamefont {S.~K.}\ \bibnamefont {Bandyopadhyay}}, \bibinfo {author} {\bibfnamefont {P.}~\bibnamefont {Sen}}, \bibinfo {author} {\bibfnamefont {U.}~\bibnamefont {Kumar}},\ and\ \bibinfo {author} {\bibfnamefont {T.~P.}\ \bibnamefont {Sinha}},\ }\bibfield  {title} {\bibinfo {title} {{Electronic structure of ordered double perovskite ${B}a_{2}{C}o{WO}_{6}$}},\ }\href {https://doi.org/10.1063/1.4872887} {\bibfield  {journal} {\bibinfo  {journal} {AIP Conference Proceedings}\ }\textbf {\bibinfo {volume} {1591}},\ \bibinfo {pages} {1155} (\bibinfo {year} {2014})}\BibitemShut {NoStop}%
\bibitem [{\citenamefont {Chatterjee}\ \emph {et~al.}(2022)\citenamefont {Chatterjee}, \citenamefont {Sau}, \citenamefont {Ghosh}, \citenamefont {Samanta}, \citenamefont {Ghosh}, \citenamefont {Kumar},\ and\ \citenamefont {Mandal}}]{chatterjee}%
  \BibitemOpen
  \bibfield  {author} {\bibinfo {author} {\bibfnamefont {S.}~\bibnamefont {Chatterjee}}, \bibinfo {author} {\bibfnamefont {J.}~\bibnamefont {Sau}}, \bibinfo {author} {\bibfnamefont {S.}~\bibnamefont {Ghosh}}, \bibinfo {author} {\bibfnamefont {S.}~\bibnamefont {Samanta}}, \bibinfo {author} {\bibfnamefont {B.}~\bibnamefont {Ghosh}}, \bibinfo {author} {\bibfnamefont {M.}~\bibnamefont {Kumar}},\ and\ \bibinfo {author} {\bibfnamefont {K.}~\bibnamefont {Mandal}},\ }\bibfield  {title} {\bibinfo {title} {Anomalous hall effect in topological weyl and nodal-line semimetal heusler compound ${C}o_{2}{V}{Al}$},\ }\href@noop {} {\bibfield  {journal} {\bibinfo  {journal} {Journal of Physics: Condensed Matter}\ }\textbf {\bibinfo {volume} {35}},\ \bibinfo {pages} {035601} (\bibinfo {year} {2022})}\BibitemShut {NoStop}%
\bibitem [{\citenamefont {Mukherjee}\ \emph {et~al.}(2014)\citenamefont {Mukherjee}, \citenamefont {Patel}, \citenamefont {Dong}, \citenamefont {Johnston}, \citenamefont {Moreo},\ and\ \citenamefont {Dagotto}}]{AM2}%
  \BibitemOpen
  \bibfield  {author} {\bibinfo {author} {\bibfnamefont {A.}~\bibnamefont {Mukherjee}}, \bibinfo {author} {\bibfnamefont {N.~D.}\ \bibnamefont {Patel}}, \bibinfo {author} {\bibfnamefont {S.}~\bibnamefont {Dong}}, \bibinfo {author} {\bibfnamefont {S.}~\bibnamefont {Johnston}}, \bibinfo {author} {\bibfnamefont {A.}~\bibnamefont {Moreo}},\ and\ \bibinfo {author} {\bibfnamefont {E.}~\bibnamefont {Dagotto}},\ }\bibfield  {title} {\bibinfo {title} {Testing the monte carlo--mean field approximation in the one-band hubbard model},\ }\href {https://doi.org/10.1103/PhysRevB.90.205133} {\bibfield  {journal} {\bibinfo  {journal} {Phys. Rev. B}\ }\textbf {\bibinfo {volume} {90}},\ \bibinfo {pages} {205133} (\bibinfo {year} {2014})}\BibitemShut {NoStop}%
\bibitem [{\citenamefont {Mukherjee}\ \emph {et~al.}(2016)\citenamefont {Mukherjee}, \citenamefont {Patel}, \citenamefont {Moreo},\ and\ \citenamefont {Dagotto}}]{AM3}%
  \BibitemOpen
  \bibfield  {author} {\bibinfo {author} {\bibfnamefont {A.}~\bibnamefont {Mukherjee}}, \bibinfo {author} {\bibfnamefont {N.~D.}\ \bibnamefont {Patel}}, \bibinfo {author} {\bibfnamefont {A.}~\bibnamefont {Moreo}},\ and\ \bibinfo {author} {\bibfnamefont {E.}~\bibnamefont {Dagotto}},\ }\bibfield  {title} {\bibinfo {title} {Orbital selective directional conductor in the two-orbital hubbard model},\ }\href {https://doi.org/10.1103/PhysRevB.93.085144} {\bibfield  {journal} {\bibinfo  {journal} {Phys. Rev. B}\ }\textbf {\bibinfo {volume} {93}},\ \bibinfo {pages} {085144} (\bibinfo {year} {2016})}\BibitemShut {NoStop}%
\bibitem [{\citenamefont {Ishizuka}\ and\ \citenamefont {Nagaosa}(2018)}]{tex-nagaosa}%
  \BibitemOpen
  \bibfield  {author} {\bibinfo {author} {\bibfnamefont {H.}~\bibnamefont {Ishizuka}}\ and\ \bibinfo {author} {\bibfnamefont {N.}~\bibnamefont {Nagaosa}},\ }\bibfield  {title} {\bibinfo {title} {Spin chirality induced skew scattering and anomalous hall effect in chiral magnets},\ }\href {https://doi.org/10.1126/sciadv.aap9962} {\bibfield  {journal} {\bibinfo  {journal} {Science Advances}\ }\textbf {\bibinfo {volume} {4}},\ \bibinfo {pages} {eaap9962} (\bibinfo {year} {2018})}\BibitemShut {NoStop}%
\bibitem [{\citenamefont {Wang}\ \emph {et~al.}(2019)\citenamefont {Wang}, \citenamefont {Daniels}, \citenamefont {Liao}, \citenamefont {Zhao}, \citenamefont {Wang}, \citenamefont {Koster}, \citenamefont {Rijnders}, \citenamefont {Chang}, \citenamefont {Xiao},\ and\ \citenamefont {Wu}}]{tex-ahc-sq}%
  \BibitemOpen
  \bibfield  {author} {\bibinfo {author} {\bibfnamefont {W.}~\bibnamefont {Wang}}, \bibinfo {author} {\bibfnamefont {M.~W.}\ \bibnamefont {Daniels}}, \bibinfo {author} {\bibfnamefont {Z.}~\bibnamefont {Liao}}, \bibinfo {author} {\bibfnamefont {Y.}~\bibnamefont {Zhao}}, \bibinfo {author} {\bibfnamefont {J.}~\bibnamefont {Wang}}, \bibinfo {author} {\bibfnamefont {G.}~\bibnamefont {Koster}}, \bibinfo {author} {\bibfnamefont {G.}~\bibnamefont {Rijnders}}, \bibinfo {author} {\bibfnamefont {C.-Z.}\ \bibnamefont {Chang}}, \bibinfo {author} {\bibfnamefont {D.}~\bibnamefont {Xiao}},\ and\ \bibinfo {author} {\bibfnamefont {W.}~\bibnamefont {Wu}},\ }\bibfield  {title} {\bibinfo {title} {Spin chirality fluctuation in two-dimensional ferromagnets with perpendicular magnetic anisotropy},\ }\href {https://doi.org/10.1038/s41563-019-0454-9} {\bibfield  {journal} {\bibinfo  {journal} {Nature Materials}\ }\textbf {\bibinfo {volume} {18}},\ \bibinfo {pages} {1054} (\bibinfo {year} {2019})}\BibitemShut {NoStop}%
\bibitem [{\citenamefont {Machida}\ \emph {et~al.}(2010)\citenamefont {Machida}, \citenamefont {Nakatsuji}, \citenamefont {Onoda}, \citenamefont {Tayama},\ and\ \citenamefont {Sakakibara}}]{tex-ahc-1}%
  \BibitemOpen
  \bibfield  {author} {\bibinfo {author} {\bibfnamefont {Y.}~\bibnamefont {Machida}}, \bibinfo {author} {\bibfnamefont {S.}~\bibnamefont {Nakatsuji}}, \bibinfo {author} {\bibfnamefont {S.}~\bibnamefont {Onoda}}, \bibinfo {author} {\bibfnamefont {T.}~\bibnamefont {Tayama}},\ and\ \bibinfo {author} {\bibfnamefont {T.}~\bibnamefont {Sakakibara}},\ }\bibfield  {title} {\bibinfo {title} {Time-reversal symmetry breaking and spontaneous hall effect without magnetic dipole order},\ }\href {https://doi.org/10.1038/nature08680} {\bibfield  {journal} {\bibinfo  {journal} {Nature}\ }\textbf {\bibinfo {volume} {463}},\ \bibinfo {pages} {210} (\bibinfo {year} {2010})}\BibitemShut {NoStop}%
\bibitem [{\citenamefont {Nakatsuji}\ \emph {et~al.}(2015)\citenamefont {Nakatsuji}, \citenamefont {Kiyohara},\ and\ \citenamefont {Higo}}]{tex-ahc-2}%
  \BibitemOpen
  \bibfield  {author} {\bibinfo {author} {\bibfnamefont {S.}~\bibnamefont {Nakatsuji}}, \bibinfo {author} {\bibfnamefont {N.}~\bibnamefont {Kiyohara}},\ and\ \bibinfo {author} {\bibfnamefont {T.}~\bibnamefont {Higo}},\ }\bibfield  {title} {\bibinfo {title} {Large anomalous hall effect in a non-collinear antiferromagnet at room temperature},\ }\href {https://doi.org/10.1038/nature15723} {\bibfield  {journal} {\bibinfo  {journal} {Nature}\ }\textbf {\bibinfo {volume} {527}},\ \bibinfo {pages} {212} (\bibinfo {year} {2015})}\BibitemShut {NoStop}%
\bibitem [{\citenamefont {Lopez}\ \emph {et~al.}(2012)\citenamefont {Lopez}, \citenamefont {Saleta}, \citenamefont {Curiale},\ and\ \citenamefont {Sanchez}}]{bcwo-tc-exp}%
  \BibitemOpen
  \bibfield  {author} {\bibinfo {author} {\bibfnamefont {C.}~\bibnamefont {Lopez}}, \bibinfo {author} {\bibfnamefont {M.}~\bibnamefont {Saleta}}, \bibinfo {author} {\bibfnamefont {J.}~\bibnamefont {Curiale}},\ and\ \bibinfo {author} {\bibfnamefont {R.}~\bibnamefont {Sanchez}},\ }\bibfield  {title} {\bibinfo {title} {Crystal field effect on the effective magnetic moment in ${A}_{2}{Co}{W}{O}_{6}$ ({A}={Ca}, {Sr} and {Ba})},\ }\href {https://doi.org/https://doi.org/10.1016/j.materresbull.2012.02.008} {\bibfield  {journal} {\bibinfo  {journal} {Materials Research Bulletin}\ }\textbf {\bibinfo {volume} {47}},\ \bibinfo {pages} {1158} (\bibinfo {year} {2012})}\BibitemShut {NoStop}%
\end{thebibliography}%


\begin{thebibliography}{14}%
\makeatletter
\providecommand \@ifxundefined [1]{%
 \@ifx{#1\undefined}
}%
\providecommand \@ifnum [1]{%
 \ifnum #1\expandafter \@firstoftwo
 \else \expandafter \@secondoftwo
 \fi
}%
\providecommand \@ifx [1]{%
 \ifx #1\expandafter \@firstoftwo
 \else \expandafter \@secondoftwo
 \fi
}%
\providecommand \natexlab [1]{#1}%
\providecommand \enquote  [1]{``#1''}%
\providecommand \bibnamefont  [1]{#1}%
\providecommand \bibfnamefont [1]{#1}%
\providecommand \citenamefont [1]{#1}%
\providecommand \href@noop [0]{\@secondoftwo}%
\providecommand \href [0]{\begingroup \@sanitize@url \@href}%
\providecommand \@href[1]{\@@startlink{#1}\@@href}%
\providecommand \@@href[1]{\endgroup#1\@@endlink}%
\providecommand \@sanitize@url [0]{\catcode `\\12\catcode `\$12\catcode
  `\&12\catcode `\#12\catcode `\^12\catcode `\_12\catcode `\%12\relax}%
\providecommand \@@startlink[1]{}%
\providecommand \@@endlink[0]{}%
\providecommand \url  [0]{\begingroup\@sanitize@url \@url }%
\providecommand \@url [1]{\endgroup\@href {#1}{\urlprefix }}%
\providecommand \urlprefix  [0]{URL }%
\providecommand \Eprint [0]{\href }%
\providecommand \doibase [0]{https://doi.org/}%
\providecommand \selectlanguage [0]{\@gobble}%
\providecommand \bibinfo  [0]{\@secondoftwo}%
\providecommand \bibfield  [0]{\@secondoftwo}%
\providecommand \translation [1]{[#1]}%
\providecommand \BibitemOpen [0]{}%
\providecommand \bibitemStop [0]{}%
\providecommand \bibitemNoStop [0]{.\EOS\space}%
\providecommand \EOS [0]{\spacefactor3000\relax}%
\providecommand \BibitemShut  [1]{\csname bibitem#1\endcsname}%
\let\auto@bib@innerbib\@empty
\bibitem [{\citenamefont {Perdew}\ and\ \citenamefont {Wang}(1992)}]{perdew}%
  \BibitemOpen
  \bibfield  {author} {\bibinfo {author} {\bibfnamefont {J.~P.}\ \bibnamefont
  {Perdew}}\ and\ \bibinfo {author} {\bibfnamefont {Y.}~\bibnamefont {Wang}},\
  }\bibfield  {title} {\bibinfo {title} {Accurate and simple analytic
  representation of the electron-gas correlation energy},\ }\href
  {https://doi.org/10.1103/PhysRevB.45.13244} {\bibfield  {journal} {\bibinfo
  {journal} {Phys. Rev. B}\ }\textbf {\bibinfo {volume} {45}},\ \bibinfo
  {pages} {13244} (\bibinfo {year} {1992})}\BibitemShut {NoStop}%
\bibitem [{\citenamefont {Hafner}(2008)}]{hafner2008ab}%
  \BibitemOpen
  \bibfield  {author} {\bibinfo {author} {\bibfnamefont {J.}~\bibnamefont
  {Hafner}},\ }\bibfield  {title} {\bibinfo {title} {Ab-initio simulations of
  materials using vasp: Density-functional theory and beyond},\ }\href@noop {}
  {\bibfield  {journal} {\bibinfo  {journal} {Journal of computational
  chemistry}\ }\textbf {\bibinfo {volume} {29}},\ \bibinfo {pages} {2044}
  (\bibinfo {year} {2008})}\BibitemShut {NoStop}%
\bibitem [{\citenamefont {Perdew}\ \emph {et~al.}(1996)\citenamefont {Perdew},
  \citenamefont {Burke},\ and\ \citenamefont {Ernzerhof}}]{burke}%
  \BibitemOpen
  \bibfield  {author} {\bibinfo {author} {\bibfnamefont {J.~P.}\ \bibnamefont
  {Perdew}}, \bibinfo {author} {\bibfnamefont {K.}~\bibnamefont {Burke}},\ and\
  \bibinfo {author} {\bibfnamefont {M.}~\bibnamefont {Ernzerhof}},\ }\bibfield
  {title} {\bibinfo {title} {Generalized gradient approximation made simple},\
  }\href {https://doi.org/10.1103/PhysRevLett.77.3865} {\bibfield  {journal}
  {\bibinfo  {journal} {Phys. Rev. Lett.}\ }\textbf {\bibinfo {volume} {77}},\
  \bibinfo {pages} {3865} (\bibinfo {year} {1996})}\BibitemShut {NoStop}%
\bibitem [{\citenamefont {Pizzi}\ \emph {et~al.}(2020)\citenamefont {Pizzi},
  \citenamefont {Vitale}, \citenamefont {Arita}, \citenamefont {Bl{\"u}gel},
  \citenamefont {Freimuth}, \citenamefont {G{\'e}ranton}, \citenamefont
  {Gibertini}, \citenamefont {Gresch}, \citenamefont {Johnson}, \citenamefont
  {Koretsune} \emph {et~al.}}]{pizzi2020wannier90}%
  \BibitemOpen
  \bibfield  {author} {\bibinfo {author} {\bibfnamefont {G.}~\bibnamefont
  {Pizzi}}, \bibinfo {author} {\bibfnamefont {V.}~\bibnamefont {Vitale}},
  \bibinfo {author} {\bibfnamefont {R.}~\bibnamefont {Arita}}, \bibinfo
  {author} {\bibfnamefont {S.}~\bibnamefont {Bl{\"u}gel}}, \bibinfo {author}
  {\bibfnamefont {F.}~\bibnamefont {Freimuth}}, \bibinfo {author}
  {\bibfnamefont {G.}~\bibnamefont {G{\'e}ranton}}, \bibinfo {author}
  {\bibfnamefont {M.}~\bibnamefont {Gibertini}}, \bibinfo {author}
  {\bibfnamefont {D.}~\bibnamefont {Gresch}}, \bibinfo {author} {\bibfnamefont
  {C.}~\bibnamefont {Johnson}}, \bibinfo {author} {\bibfnamefont
  {T.}~\bibnamefont {Koretsune}}, \emph {et~al.},\ }\bibfield  {title}
  {\bibinfo {title} {Wannier90 as a community code: new features and
  applications},\ }\href@noop {} {\bibfield  {journal} {\bibinfo  {journal}
  {Journal of Physics: Condensed Matter}\ }\textbf {\bibinfo {volume} {32}},\
  \bibinfo {pages} {165902} (\bibinfo {year} {2020})}\BibitemShut {NoStop}%
\bibitem [{\citenamefont {Marzari}\ and\ \citenamefont
  {Vanderbilt}(1997)}]{MarzariPhysRevB.56.12847}%
  \BibitemOpen
  \bibfield  {author} {\bibinfo {author} {\bibfnamefont {N.}~\bibnamefont
  {Marzari}}\ and\ \bibinfo {author} {\bibfnamefont {D.}~\bibnamefont
  {Vanderbilt}},\ }\bibfield  {title} {\bibinfo {title} {Maximally localized
  generalized wannier functions for composite energy bands},\ }\href
  {https://doi.org/10.1103/PhysRevB.56.12847} {\bibfield  {journal} {\bibinfo
  {journal} {Phys. Rev. B}\ }\textbf {\bibinfo {volume} {56}},\ \bibinfo
  {pages} {12847} (\bibinfo {year} {1997})}\BibitemShut {NoStop}%
\bibitem [{\citenamefont {Wu}\ \emph {et~al.}(2018)\citenamefont {Wu},
  \citenamefont {Zhang}, \citenamefont {Song}, \citenamefont {Troyer},\ and\
  \citenamefont {Soluyanov}}]{wu2018wanniertools}%
  \BibitemOpen
  \bibfield  {author} {\bibinfo {author} {\bibfnamefont {Q.}~\bibnamefont
  {Wu}}, \bibinfo {author} {\bibfnamefont {S.}~\bibnamefont {Zhang}}, \bibinfo
  {author} {\bibfnamefont {H.-F.}\ \bibnamefont {Song}}, \bibinfo {author}
  {\bibfnamefont {M.}~\bibnamefont {Troyer}},\ and\ \bibinfo {author}
  {\bibfnamefont {A.~A.}\ \bibnamefont {Soluyanov}},\ }\bibfield  {title}
  {\bibinfo {title} {Wanniertools: An open-source software package for novel
  topological materials},\ }\href@noop {} {\bibfield  {journal} {\bibinfo
  {journal} {Computer Physics Communications}\ }\textbf {\bibinfo {volume}
  {224}},\ \bibinfo {pages} {405} (\bibinfo {year} {2018})}\BibitemShut
  {NoStop}%
\bibitem [{\citenamefont {Gradhand}\ \emph {et~al.}(2012)\citenamefont
  {Gradhand}, \citenamefont {Fedorov}, \citenamefont {Pientka}, \citenamefont
  {Zahn}, \citenamefont {Mertig},\ and\ \citenamefont
  {Györffy}}]{Gradhand_2012}%
  \BibitemOpen
  \bibfield  {author} {\bibinfo {author} {\bibfnamefont {M.}~\bibnamefont
  {Gradhand}}, \bibinfo {author} {\bibfnamefont {D.~V.}\ \bibnamefont
  {Fedorov}}, \bibinfo {author} {\bibfnamefont {F.}~\bibnamefont {Pientka}},
  \bibinfo {author} {\bibfnamefont {P.}~\bibnamefont {Zahn}}, \bibinfo {author}
  {\bibfnamefont {I.}~\bibnamefont {Mertig}},\ and\ \bibinfo {author}
  {\bibfnamefont {B.~L.}\ \bibnamefont {Györffy}},\ }\href
  {https://doi.org/10.1088/0953-8984/24/21/213202} {\bibfield  {journal}
  {\bibinfo  {journal} {Journal of Physics: Condensed Matter}\ }\textbf
  {\bibinfo {volume} {24}},\ \bibinfo {pages} {213202} (\bibinfo {year}
  {2012})}\BibitemShut {NoStop}%
  
\bibitem [{\citenamefont {Xiao}\ \emph {et~al.}(2010)\citenamefont {Xiao},
\citenamefont {Chang},\ and\ \citenamefont {Niu}}]{xiao2010berry}%
\BibitemOpen
\bibfield  {author} {\bibinfo {author} {\bibfnamefont {D.}~\bibnamefont
{Xiao}}, \bibinfo {author} {\bibfnamefont {M.-C.}\ \bibnamefont {Chang}},\
and\ \bibinfo {author} {\bibfnamefont {Q.}~\bibnamefont {Niu}},\ }\bibfield
{title} {\bibinfo {title} {Berry phase effects on electronic properties},\
}\href@noop {} {\bibfield  {journal} {\bibinfo  {journal} {Reviews of Modern
Physics}\ }\textbf {\bibinfo {volume} {82}},\ \bibinfo {pages} {1959}
(\bibinfo {year} {2010})}\BibitemShut {NoStop}%

  
  
\bibitem [{\citenamefont {Patel}\ \emph {et~al.}(2017)\citenamefont {Patel},
  \citenamefont {Mukherjee}, \citenamefont {Kaushal}, \citenamefont {Moreo},\
  and\ \citenamefont {Dagotto}}]{patel}%
  \BibitemOpen
  \bibfield  {author} {\bibinfo {author} {\bibfnamefont {N.~D.}\ \bibnamefont
  {Patel}}, \bibinfo {author} {\bibfnamefont {A.}~\bibnamefont {Mukherjee}},
  \bibinfo {author} {\bibfnamefont {N.}~\bibnamefont {Kaushal}}, \bibinfo
  {author} {\bibfnamefont {A.}~\bibnamefont {Moreo}},\ and\ \bibinfo {author}
  {\bibfnamefont {E.}~\bibnamefont {Dagotto}},\ }\bibfield  {title} {\bibinfo
  {title} {Non-fermi liquid behavior and continuously tunable resistivity
  exponents in the anderson-hubbard model at finite temperature},\ }\href
  {https://doi.org/10.1103/PhysRevLett.119.086601} {\bibfield  {journal}
  {\bibinfo  {journal} {Phys. Rev. Lett.}\ }\textbf {\bibinfo {volume} {119}},\
  \bibinfo {pages} {086601} (\bibinfo {year} {2017})}\BibitemShut {NoStop}%
\bibitem [{\citenamefont {Mukherjee}\ \emph {et~al.}(2014)\citenamefont
  {Mukherjee}, \citenamefont {Patel}, \citenamefont {Dong}, \citenamefont
  {Johnston}, \citenamefont {Moreo},\ and\ \citenamefont {Dagotto}}]{AM2}%
  \BibitemOpen
  \bibfield  {author} {\bibinfo {author} {\bibfnamefont {A.}~\bibnamefont
  {Mukherjee}}, \bibinfo {author} {\bibfnamefont {N.~D.}\ \bibnamefont
  {Patel}}, \bibinfo {author} {\bibfnamefont {S.}~\bibnamefont {Dong}},
  \bibinfo {author} {\bibfnamefont {S.}~\bibnamefont {Johnston}}, \bibinfo
  {author} {\bibfnamefont {A.}~\bibnamefont {Moreo}},\ and\ \bibinfo {author}
  {\bibfnamefont {E.}~\bibnamefont {Dagotto}},\ }\bibfield  {title} {\bibinfo
  {title} {Testing the monte carlo--mean field approximation in the one-band
  hubbard model},\ }\href {https://doi.org/10.1103/PhysRevB.90.205133}
  {\bibfield  {journal} {\bibinfo  {journal} {Phys. Rev. B}\ }\textbf {\bibinfo
  {volume} {90}},\ \bibinfo {pages} {205133} (\bibinfo {year}
  {2014})}\BibitemShut {NoStop}%
\bibitem [{\citenamefont {Chakraborty}\ \emph {et~al.}(2022)\citenamefont
  {Chakraborty}, \citenamefont {Mukherjee},\ and\ \citenamefont
  {Pradhan}}]{chakraborty}%
  \BibitemOpen
  \bibfield  {author} {\bibinfo {author} {\bibfnamefont {S.}~\bibnamefont
  {Chakraborty}}, \bibinfo {author} {\bibfnamefont {A.}~\bibnamefont
  {Mukherjee}},\ and\ \bibinfo {author} {\bibfnamefont {K.}~\bibnamefont
  {Pradhan}},\ }\bibfield  {title} {\bibinfo {title} {Antiferromagnetism beyond
  the classical percolation threshold in the diluted half-filled one-band
  hubbard model in three dimensions},\ }\href
  {https://doi.org/10.1103/PhysRevB.106.075146} {\bibfield  {journal} {\bibinfo
   {journal} {Phys. Rev. B}\ }\textbf {\bibinfo {volume} {106}},\ \bibinfo
  {pages} {075146} (\bibinfo {year} {2022})}\BibitemShut {NoStop}%
\bibitem [{\citenamefont {Mukherjee}\ \emph {et~al.}(2016)\citenamefont
  {Mukherjee}, \citenamefont {Patel}, \citenamefont {Moreo},\ and\
  \citenamefont {Dagotto}}]{AM3}%
  \BibitemOpen
  \bibfield  {author} {\bibinfo {author} {\bibfnamefont {A.}~\bibnamefont
  {Mukherjee}}, \bibinfo {author} {\bibfnamefont {N.~D.}\ \bibnamefont
  {Patel}}, \bibinfo {author} {\bibfnamefont {A.}~\bibnamefont {Moreo}},\ and\
  \bibinfo {author} {\bibfnamefont {E.}~\bibnamefont {Dagotto}},\ }\bibfield
  {title} {\bibinfo {title} {Orbital selective directional conductor in the
  two-orbital hubbard model},\ }\href
  {https://doi.org/10.1103/PhysRevB.93.085144} {\bibfield  {journal} {\bibinfo
  {journal} {Phys. Rev. B}\ }\textbf {\bibinfo {volume} {93}},\ \bibinfo
  {pages} {085144} (\bibinfo {year} {2016})}\BibitemShut {NoStop}%
\bibitem [{\citenamefont {Luo}\ \emph {et~al.}(2011)\citenamefont {Luo},
  \citenamefont {Yao}, \citenamefont {Moreo},\ and\ \citenamefont
  {Dagotto}}]{luo}%
  \BibitemOpen
  \bibfield  {author} {\bibinfo {author} {\bibfnamefont {Q.}~\bibnamefont
  {Luo}}, \bibinfo {author} {\bibfnamefont {D.-X.}\ \bibnamefont {Yao}},
  \bibinfo {author} {\bibfnamefont {A.}~\bibnamefont {Moreo}},\ and\ \bibinfo
  {author} {\bibfnamefont {E.}~\bibnamefont {Dagotto}},\ }\bibfield  {title}
  {\bibinfo {title} {Charge stripes in the two-orbital hubbard model for iron
  pnictides},\ }\href {https://doi.org/10.1103/PhysRevB.83.174513} {\bibfield
  {journal} {\bibinfo  {journal} {Phys. Rev. B}\ }\textbf {\bibinfo {volume}
  {83}},\ \bibinfo {pages} {174513} (\bibinfo {year} {2011})}\BibitemShut
  {NoStop}%
\bibitem [{\citenamefont {Kumar}\ and\ \citenamefont {Majumdar}(2006)}]{kumar}%
  \BibitemOpen
  \bibfield  {author} {\bibinfo {author} {\bibfnamefont {S.}~\bibnamefont
  {Kumar}}\ and\ \bibinfo {author} {\bibfnamefont {P.}~\bibnamefont
  {Majumdar}},\ }\bibfield  {title} {\bibinfo {title} {A travelling cluster
  approximation for lattice fermions strongly coupled to classical degrees of
  freedom},\ }\href {https://doi.org/10.1140/epjb/e2006-00173-2} {\bibfield
  {journal} {\bibinfo  {journal} {The European Physical Journal B - Condensed
  Matter and Complex Systems}\ }\textbf {\bibinfo {volume} {50}},\ \bibinfo
  {pages} {571} (\bibinfo {year} {2006})}\BibitemShut {NoStop}%
\end{thebibliography}%

\end{document}


\title{Supplemental Material: Emergent scalar-chirality \& colossal transverse-magnetoresponse in strongly correlated nodal-line half-metal.} 

\author{Jyotirmoy Sau}
\thanks{\textcolor{magenta}{These authors contributed equally to this work.}}
\affiliation{Department of Condensed Matter and Materials Physics,
S. N. Bose National Centre for Basic Sciences, JD Block, Sector III, Salt Lake, Kolkata 700106, India}

\author{Sourav Chakraborty}
\thanks{\textcolor{magenta}{These authors contributed equally to this work.}}
\affiliation{Department of Condensed Matter and Materials Physics,
S. N. Bose National Centre for Basic Sciences, JD Block, Sector III, Salt Lake, Kolkata 700106, India}

\author{Sourabh Saha}
\affiliation{Department of Condensed Matter and Materials Physics,
S. N. Bose National Centre for Basic Sciences, JD Block, Sector III, Salt Lake, Kolkata 700106, India}

\author{Kalpataru Pradhan}
\email{kalpataru.pradhan@saha.ac.in}
\affiliation{Theory Division, Saha Institute of Nuclear Physics,
  A CI of Homi Bhabha National Institute, Kolkata-700064, India}

\author{Anamitra Mukherjee}
\email{anamitra@niser.ac.in}
\affiliation{School of Physical Sciences, National Institute of Science Education and Research,
a CI of Homi Bhabha National Institute, Jatni 752050, India}

\author{Manoranjan Kumar}
\email{manoranjan.kumar@bose.res.in}
\affiliation{Department of Condensed Matter and Materials Physics,
S. N. Bose National Centre for Basic Sciences, JD Block, Sector III, Salt Lake, Kolkata 700106, India}

\begin{abstract}
We here provide supplemental explanations and data on the following topics in relation to the main text: I. Details of the first principles methodologies and Wannierization leading to down-folded Hamiltonian. II. Details of the down-folded Hamiltonian and the anomalous Hall conductivity within first-principles calculations.
 III. contains the details of the multi-orbital and spin-orbit Hamiltonian. IV discussed the exact diagonalization results. In V we discuss the semi-classical Monte-Carlo approach and the Kubo-Greenwood formalism for conductivity is discussed in VI. In VII we provide results on the field dependence of the longitudinal conductivity as a function of temperature.
\end{abstract}
\maketitle









\section{First-principles approach}
\subsection{\label{sec2} Details of first-principles method }

Ab initio:  We have also started the DFT computation for the material $A_{2}BB'O_{6}$ ($A = Ba; B = Co; B' = W$) by adding a coulomb repulsion for the treatment of the exchange and correlation effects for Co (3d) and W (5d) electrons. The Hubbard technique or GGA+U \cite{perdew}has been applied. For the 3d and 5d states, respectively, the onsite Coulomb energy (U) parameters are used  ($U_{eff}$ = 4.0 eV and ($U_{eff}$ = 1.0 eV) respectively. In addition, we investigated the electronic band structures and transport properties of $A_{2}BB'O_{6}$ using first-principles electronic structure simulations.  To depict the ion-electron interactions, the projected augmented wave approach is used. The computations are completed using the Vienna ab initio simulation package (VASP) \cite{hafner2008ab}. For the electronic exchange-correlation functional, Perdew-Burke-Ernzerhof (PBE)  \cite{burke} generalised gradient approximation (GGA) \cite{perdew} of perdew-burke type is utilised. For the plane-wave basis, a kinetic energy cutoff of 500 eV is used. The electronic integral over BZ is estimated by the Gaussian smearing method with a width of 0.01 eV, and the BZ sampling is carried out on a 16 $\times$ 16 $\times$ 16 Monkhorst-pack k-point mesh with a $\Gamma$-centered grid. Starting from the plane wave basis state, Energy Gap, BC, and AHC are estimated using Wannier90 \cite{pizzi2020wannier90,MarzariPhysRevB.56.12847} and Wanniertools \cite{wu2018wanniertools}.

The positions of Ba(A), Co, Fe($B$), W, Re($B'$), O$_1$ \& O$_2$ are 4c (0,0.5,0.25), 2b(0,0,0.5), 2a(0,0,0), 4e(x,y,0) $\&$ (0,0,z). The structure adopts a Face-centered cubic crystal structure with space group \textit{Fm$\bar{3}$m} (space group no. 225) and O$_{h}^5$ octahedral symmetry with a lattice parameter of a = 7.890 \r{A}. This is consistent with the optimized structure obtained using ab initio calculations in the same space group. 

\subsection{\label{SM4} Characteristics of the nodal line (NL) in the presence and absence of spin-orbit coupling (SOC).}
 
We've illustrated the Zoom version of NL crossing in Figure (a) without SOC, where red and blue colors represent distinct orthogonal d-orbitals. With the presence of SOC, this NL becomes gapped out with band inversion, depicted in Figure (b). In the 2D K-plane, the NL is represented by the white region in Figure (c), and corresponding to this gapped NL, a large berry curvature is observed, indicated by the red line in Figure (d).

\begin{figure}[h!]
	\centering
	\includegraphics[width=0.8\textwidth]{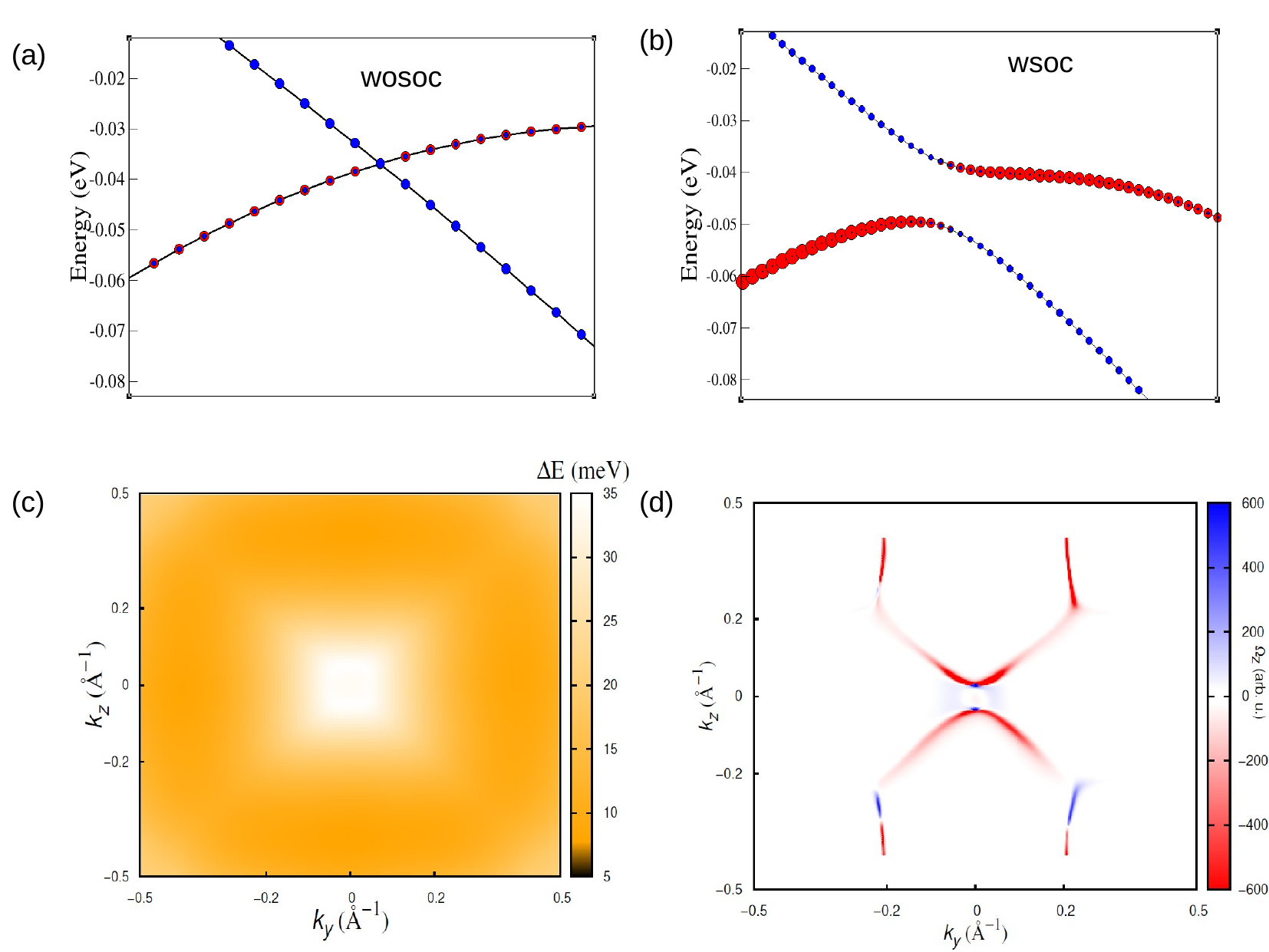}
	\caption{(a) The zoomed version of the nodal crossing point without SOC (b) The zoomed version of the nodal crossing point in the presence of SOC which shows the band inversion (c)The nodal line is gapped out in the presence of SOC in the k$_y$-k$_z$ plane, and the gap is represented by a white line. (d) The Berry curvature distribution is displayed in the k$_y$-k$_z$ plane along the gapped nodal line.}
\label{fig:S2}
\end{figure}

\section{Down-folded Hopping Matrix \& Ab-initio calculation of AHC:}
We derive Wannier functions from the DFT band structure using the WANNIER90 package \cite{pizzi2020wannier90,MarzariPhysRevB.56.12847}. The initial projections are chosen as the s orbital of Ba, d orbitals of Co and W, and p orbital of O. The Wannier functions yield the following hopping matrix, which we used in our calculation:
\begin{equation}
\begin{pmatrix}
0.2 & 0.08 & 0.08  \\
0.08 & 0.2 & 0.08  \\
0.08 & 0.08 & 0.2  \\
\end{pmatrix}    
\end{equation}

We formed this hopping matrix using the basis of \{$d^{\dagger}_{xz},d^{\dagger}_{yz},d^{\dagger}_{xy}$\} in a calculation \textit{without SOC}. Our Wannier calculations indicate that $e_g$ orbitals do not participate in the hopping process. So, we're disregarding the hopping.
The intrinsic AHC can be evaluated in the linear response theory of Kubo formalism\cite{Gradhand_2012} and AHC in xy plane can be expressed as  

\begin{equation}
 \sigma_{xy} = -{\frac{e^2}{\hbar} \int\frac{d^{3}\textit{k}}{(2\pi)^3}\sum_{n}\Omega^z_{n}(\textit{k})f_n(\textit{k})}  \end{equation}
 $\Omega^{z}_n$  is Berry curvature and it can be written as\cite{xiao2010berry}
 \begin{eqnarray}
\Omega^z_{n} = -2i \sum_{m \neq n} \frac{{\langle \psi_{n\textit{k}}|v_x|\psi_{m\textit{k}}\rangle} {\langle \psi_{m\textit{k}}|v_y|\psi_{n\textit{k}}\rangle}} {[E_{m}(\textit{k}) - E_{n}(\textit{k})]^2}
\end{eqnarray},

where $f_{n}(\textit{k})$ is the Fermi-Dirac distribution function, n is an index of the occupied bands, $E_{n}(\textit{k})$ is the eigenvalue of the $n^{\text{th}}$ eigen-state $\psi_{n}(\textit{k})$,$v_i$ = $\frac{1}{\hbar}\frac{\partial H(\textit{k})}{\partial \textit{k}_i }$ is the velocity operator along the i (i = x, y, z)

\section{A. Multi-orbital Hubbard model}

The multi-band Hubbard model can be written in three parts : 

\begin{eqnarray}
H = \sum_{i} H_{i} + \sum_{i}H_{i}^{soc}+\sum_{<ij>}H_{ij}
\label{Sq4}
\end{eqnarray} 

The first single site term, $H_{i}$ contains information of intra-band and inter-band Coulomb repulsion and inter-band Hund's coupling as well as chemical potential of different interacting orbitals. The second term, $H_{i}^{soc}$ represents the spin-orbit coupling between orbitals of a specific site. The third term, $H_{ij}$ denotes the electron hopping between the neighbouring Co and W orbitals.

\begin{eqnarray}
H_{i} =\sum_{\alpha}(\Delta_{i,\alpha}n_{i,\alpha}+ U_{i} 
 n_{i\uparrow \alpha}n_{i\downarrow \alpha}) +  \sum_{\alpha < \beta} (U_{i}' - \frac{J_{H,i}}{2}) n_{i \alpha}n_{i \beta}
 \label{Sq5}\\\nonumber
- 2 J_{H,i} \sum_{ \alpha < \beta} \mathbf{S}_{i \alpha}\cdot\mathbf{S}_{i \beta} 
+ J_{H,i}' \sum_{\alpha < \beta} (c^{\dagger}_{i\uparrow \alpha}c^{\dagger}_{i\downarrow \alpha}c_{i\downarrow \beta}c_{i\uparrow \beta} + h.c. )
\end{eqnarray}

 here $\Delta_{i,\alpha}$ is site energy of i$^{th}$ site and $\alpha$ orbitals and $U_{i,\alpha}$ is Hubbard interaction of site $i$ and $\alpha$-orbital. $(U_{i}' - \frac{J_{i,H}}{2})$ is coulomb interaction between the electrons on different orbitals of the same site $i$. $J_{i,H}$ and $J'_{i,H}$ denote the strengths of Hund's coupling and pair hopping, respectively, between different orbitals at the $i^{th}$ site. Due to symmetry, we set $J'_{i,H}$=$J_{i,H}$. In our calculation both Co  and W are have multi-band d-orbitals. 

The spin-orbital interaction Hamiltonian can be written as 
\begin{eqnarray}
    H_{i}^{soc}=\mathbf{\lambda} \mathbf{L}_{i} . \mathbf{S}_{i} \equiv \sum_{\alpha,\beta}\lambda_{\alpha,\beta}c^{\dagger}_{i\alpha}c_{i\beta},
\end{eqnarray}
where $\lambda_{\alpha,\beta}$ is exchange strength for $\alpha$ and $\beta$ d-orbitals and the exchange term generates mixing among spins and orbitals. Using the matrix elements of atomic orbital angular momentum and Pauli spin matrices the SOC interaction can be written in the basis, $\psi^{\dagger} = \{ d^{\dagger}_{xz \uparrow}, d^{\dagger}_{yz \uparrow}, d^{\dagger}_{xy \downarrow}, d^{\dagger}_{3z^{2} - r^{2} \downarrow}, d^{\dagger}_{x^{2} - y^{2} \downarrow}, d^{\dagger}_{xz \downarrow}, d^{\dagger}_{yz \downarrow}, d^{\dagger}_{xy \uparrow}, d^{\dagger}_{3z^{2} - r^{2} \uparrow}, d^{\dagger}_{x^{2} - y^{2} \uparrow}\}$ - \\

\begin{eqnarray}
H_{SOC} = \psi^{\dagger}A \psi \nonumber
\end{eqnarray}

\begin{eqnarray}
A= \lambda
\begin{pmatrix}
0 & -i & i & \sqrt{3} & -1 & 0 & 0 & 0 & 0 & 0 \\
i & 0 & -1 & -i\sqrt{3} & -i & 0 & 0 & 0 & 0 & 0 \\
-i & -1 & 0 & 0 & -2i & 0 & 0 & 0 & 0 & 0 \\
\sqrt{3} & i\sqrt{3} & 0 & 0 & 0 & 0 & 0 & 0 & 0 & 0 \\
-1 & i & 2i & 0 & 0 & 0 & 0 & 0 & 0 & 0 \\
0 & 0 & 0 & 0 & 0 & 0 & i & i & -\sqrt{3} & 1 \\
0 & 0 & 0 & 0 & 0 & -i & 0 & 1 & -i\sqrt{3} & -i \\
0 & 0 & 0 & 0 & 0 & -i & 1 & 0 & 0 & 2i \\
0 & 0 & 0 & 0 & 0 & -\sqrt{3} & i\sqrt{3} & 0 & 0 & 0 \\
0 & 0 & 0 & 0 & 0 & 1 & -i & 2i & 0 & 0 \nonumber
\end{pmatrix}
\end{eqnarray}

The inter site  hopping of electrons term $H_{ij}$ can be written as 
\begin{eqnarray}
H_{ij}=- \sum_{ \sigma, \alpha \beta} t_{\alpha, \beta} (c^{\dagger}_{i\sigma \alpha}c_{j\sigma \beta} + H.c.),
\end{eqnarray}
where $t_{\alpha \beta}$ is the hopping strength between the orbital $\alpha$ of $i$-th Co site and orbital $\beta$ of $j$-th W site. $c^{\dagger}_{i\sigma \alpha}$ and $c_{j\sigma \beta}$ are the creation operator of $\sigma$ spin, at $\alpha$ orbital of $i^{th}$ Co and  annihilation operator at $\beta$-orbital of $j^{th}$ W. The electronic hopping is restricted to the three $t_{2g}$ orbitals of Co and W as DFT calculations suggest and matrix of the  $t_{ \alpha,\beta}$  elements are shown in Sec. II above.  
\section{Details of Exact diagonalization calculations} 
\subsection{Ground state electronic configuration of a single Co  atom }

\begin{figure}[h]
	\centering
	\includegraphics[width=0.8\textwidth]{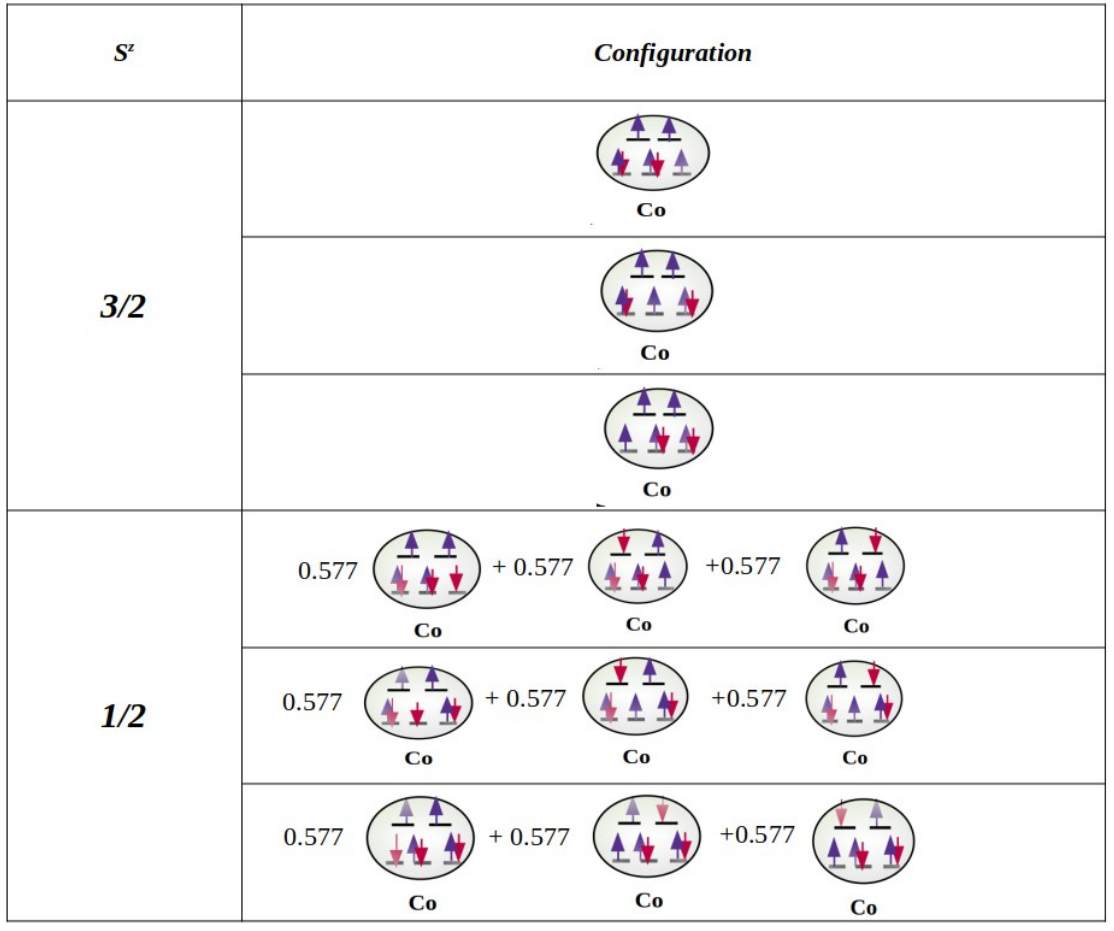}
	\caption{Ground state many body configurations of a Co atom with five d orbitals and 7 electrons for $S^z=$3/2 and 1/2.}
    \label{S1}
\end{figure}

We use the multi-band Hubbard model as described by  Eq.\ref{Sq5}  to calculate the ground state (gs) energy and wave function for a single-site Co atom. We first consider all five d-orbitals of a  Co atom which contains 7 electrons and the splitting of t$_{2g}$ and e$_{g}$ energy levels is in the order of $\Delta$ = 1.2 eV. For the given parameter the gs spin is  S=3/2 and it is 12-fold degenerate. There are (2S+1)=4 fold spin degeneracy and each of these states are three fold degenerate due to orbital degrees of freedom.  The detail of many body configurations with their coefficient is given for $S^z$=3/2 and 1/2 in Fig. \ref{S1}.

\begin{figure}[h]
	\centering
	\includegraphics[width=0.6\textwidth]{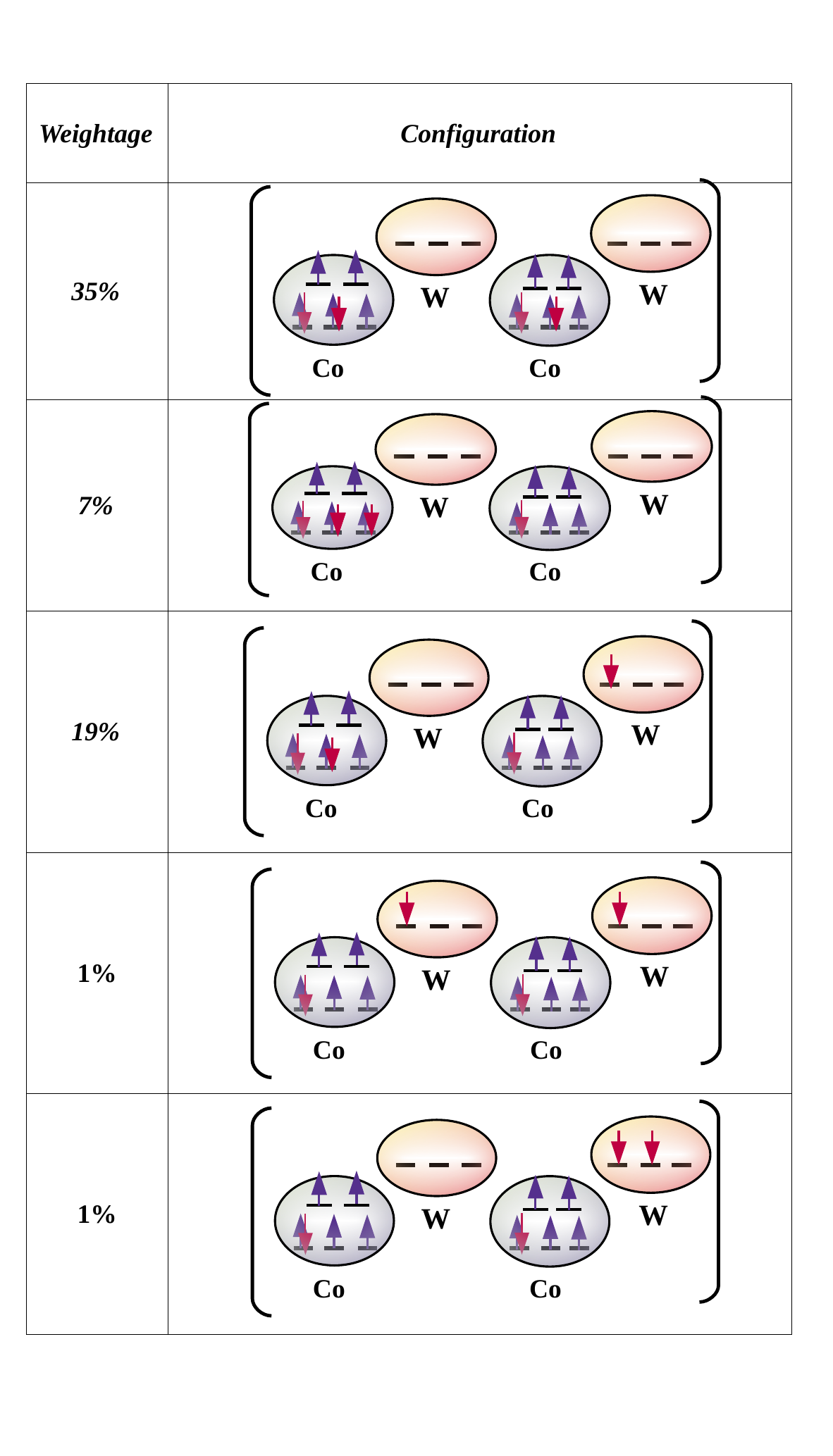}
	\caption{Five dominant gs electronic configurations and their weightage of the Co -W chain.}
\label{S2}
\end{figure}

\subsection{\label{SM2}Ground state electronic configurations of Co-W chain}

Let us take four site chains of two Co and two W atoms shown in Fig.\ref{S2} and we solve the full Hamiltonian given in Eq. \ref{Sq4} without the SOC term to model this chain. The gs is a linear combination of many state electronic configurations, and we show five dominant configurations.  The most dominant configuration contributes 35 \% of the wave function and has six possible rearrangements. It suggests two doubly occupied and one singly occupied $t_{2g}$ orbitals and two singly occupied $e_{g}$ orbitals of Co. The singly occupied orbitals favor up spin alignment due to Hund's coupling and W orbitals are vacant, which is expected from DFT and mean field Monte Carlo (MFMC) calculations.  The second configuration contributes only 7 \% and has 6 possible rearrangements, featuring one of the Co atom has all three doubly occupied $t_{2g}$ orbitals and  both  $e_{g}$ orbitals are singly occupied, while the second  Co site has one doubly and two singly occupied $t_{2g}$ orbital and two singly occupied $e_{g}$ orbitals. Here also W orbitals are vacant.

 In the third configuration( contributes 19 \% and has 24 fold possible rearrangements), we observed one down spin electron hops from one of the doubly occupied $t_{2g}$ orbitals of Co to one of the orbitals of W because of favorable Hund's coupling at Co sites, while the remaining spin configuration remains same like the first configuration.  The fourth and fifth configurations ( both of these contribute only 1\% and have 6 possible rearrangements ) depict the possibility of two down spin electrons in W sites which hop from $t_{2g}$ orbitals of Co to W orbitals to maximize the Hund's coupling at Co sites. 
  
  In Fig.3 in the main text, the gs charge density, $<n_{t_{2g}}>_{Co}$ and spin density, $< S^z_{t_{2g}}>_{Co}$ of $t_{2g}$ orbitals of Co atoms are 1.543 and 0.19, whereas for Co $e_g$,  $<n_{e_g}>_{Co}$ and $< S^z_{e_{g}}>_{Co}$ are 1 and 0.5 respectively. The $<n_{t_{2g}}>_{W}$ and $< S^z_{t_{2g}}>_{W}$ of W are 0.124 and  -0.022 respectively. The spin densities at Co sites in ED calculations are relatively higher than that obtained from our MFMC method, and this may be because of thermal fluctuations and charge density fluctuations in larger systems used in the MFMC calculations. We outline the MFMC method in the next section.

\label{S3}

\section{Semi-classical Monte-Carlo}

The semi-classical Monte-Carlo scheme rests upon admitting thermal fluctuations on Auxiliary fields (Aux. F.) obtained after the many-body Hamiltonian is treated in the saddle-point approximation. It misses out on the temporal fluctuations of the Aux. F., but by admitting thermal fluctuations within a Classical Monte-Carlo coupled with exact diagonalization of the saddle point Hamiltonian, several physical aspects of the many-body problem at finite temperature are captured. These results are well beyond finite temperature mean-field theory \cite{patel, AM2, chakraborty}. The saddle point Hamiltonian can be obtained by performing a Hubbard-Stratonovich decomposition starting from the full partition function by introducing Aux. F. \cite{AM2} and then treating the Aux. F as static variables. This approximation (of static Aux. F.) allows one to extract an effective Hamiltonian with fermions coupled to classical Aux. F.  to be annealed within a classical Monte-Carlo coupled with exact diagonalization. The same Hamiltonian can also be obtained by calculating the mean field decomposition and treating the mean-field variables as classical Aux. F. variables\cite{AM3}. Here, we follow the latter approach. The on-site Hamiltonian in Eq. 5 can be decomposed within a mean-field theory to obtain the following local effective Hamiltonian. In the mean-field Hamiltonian, we have vector Aux. F. \textbf{m}$_{i,\alpha}$, charge density Aux. F. $\{<n_{i,\alpha}>\}$ and Aux. F. corresponding to cross-orbital two-fermion expectations. The mean-field decomposition is identical to previous literature on the five-orbital Hubbard model \cite{luo}. The effective Hamiltonian can be written as,



\begin{eqnarray}
H_{i}^{eff} = \sum_{\alpha}\Delta_{i,\alpha}n_{i,\alpha} + 
+ \frac{U_{i}}{2} (<n_{i,\alpha}> n_{i,\alpha} - \textbf{m}_{i,\alpha} . \sigma_{i,\alpha}) + \sum_{\alpha \neq \beta} (U_{i}' - \frac{J_{H,i}}{2}) \hspace{0.1cm} n_{i, \alpha}<n_{i, \beta}> -\frac{J_{H,i}}{2} \hspace{0.1cm} \textbf{m}_{i,\alpha} . \sigma_{i,\beta} -\mu n_i \nonumber \\
  + \sum_{\alpha} \frac{U_{i}}{4}(\textbf{m}_{i,\alpha}^2 - {<n_{i,\alpha}>}^2) -  \sum_{\alpha \neq \beta} \frac{1}{2} (U_{i}' - \frac{J_{H,i}}{2}) \hspace{0.1cm} <n_{i, \alpha}><n_{i, \beta}> +  \frac{J_{H,i}}{4} \hspace{0.1cm} \textbf{m}_{i,\alpha} . \textbf{m}_{i,\beta}  \nonumber \\
- \sum_{\alpha \neq \beta, \sigma, \sigma'} (U_{i}' - \frac{J_{H,i}}{2}) [c^{\dagger}_{i, \alpha, \sigma}c_{i, \beta, \sigma'} <c^{\dagger}_{i, \beta, \sigma'}c_{i, \alpha, \sigma}> - \frac{1}{2} <c^{\dagger}_{i, \alpha, \sigma}c_{i, \beta, \sigma'}> <c^{\dagger}_{i, \beta, \sigma'}c_{i, \alpha, \sigma}> ] \nonumber \\
+ \sum_{\alpha \neq \beta, \sigma} J_{H,i} [c^{\dagger}_{i, \alpha, \sigma}c_{i, \beta, \sigma}<c^{\dagger}_{i, \beta, -\sigma}c_{i, \alpha, -\sigma}> -\frac{1}{2} <c^{\dagger}_{i, \alpha, \sigma}c_{i, \beta, \sigma}><c^{\dagger}_{i, \beta, -\sigma}c_{i, \alpha, -\sigma}>] \nonumber \\
+ \sum_{\alpha \neq \beta, \sigma, \sigma'} \frac{J_{H,i}}{2} [ (-1)^{(1 - \delta_{\sigma, \sigma'})}c^{\dagger}_{i, \alpha, \sigma}c_{i, \beta, \sigma'} <c^{\dagger}_{i, \beta, \sigma'}c_{i, \alpha, \sigma}> - \frac{1}{2} <c^{\dagger}_{i, \alpha, \sigma}c_{i, \beta, \sigma'}> <c^{\dagger}_{i, \beta, \sigma'}c_{i, \alpha, \sigma}> ] \nonumber \\
+ \sum_{\alpha \neq \beta, \sigma, \sigma'} J_{H,i}' [ (-1)^{(1 - \delta_{\sigma, \sigma'})}c^{\dagger}_{i, \alpha, \sigma}c_{i, \beta, \sigma'} <c^{\dagger}_{i, \alpha, -\sigma}c_{i, \beta, -\sigma'}> - \frac{1}{2} <c^{\dagger}_{i, \alpha, \sigma}c_{i, \beta, \sigma'}> <c^{\dagger}_{i, \alpha, -\sigma}c_{i, \beta, -\sigma'}> ] \nonumber \\
\end{eqnarray}

We start the Monte-Carlo process with randomly chosen spin auxiliary fields \{\textbf{m}$_{i,\alpha}$\} and a uniform value for the charge density $\{<n_{i,\alpha}>\}$ and cross-orbital two-fermion expectations. We construct the Hamiltonian, including the kinetic and SOC terms, and diagonalize it. We update the Aux. F.  \{\textbf{m}$_{i,\alpha}$\} at a site (for each orbital separately) and accept the update within a Metropolis scheme. We perform the updating process for all sites separately, constituting a system sweep. After every five system sweeps, the charge density and cross-orbital fermion expectations are calculated self-consistently and fed back into the Monte Carlo scheme.

We perform these calculations within a simple cubic lattice where neighboring sites correspond to Co and W atoms, and each site has five orbitals. Therefore, an $N^3$ system signifies a $2\times5\times N^{3}$ dimensional calculation where the multiplication of 2 is due to the spin indices. Due to the enormous computational cost, we perform our calculations for the N=6 system using a traveling cluster approximation \cite{kumar}, employing a $4^3$ traveling cluster. We use 2000 Monte Carlo system sweeps for each temperature, where the first 1000 sweeps are used to thermalize the system, and the rest are used to compute the thermally averaged observables, discarding ten sweeps in between to avoid self-correlation. 

\section{\label{SM1}Brief description of conductivity calculation}

We calculate the longitudinal and transverse conductivity using the Kubo-Greenwood formula. The longitudinal part of optical conductivity can be expressed as - 

\begin{eqnarray}
\sigma(\omega) = \frac{A}{N} \sum_{\alpha, \beta} (n_{\alpha} - n_{\beta})\frac{|f^{x}_{\alpha \beta}|^{2}}{\epsilon_{\beta} - \epsilon_{\alpha}} \delta[\omega - (\epsilon_{\beta} - \epsilon_{\alpha})] \nonumber
\end{eqnarray}

Here the constant term $A = \frac{\pi e^{2}}{\hbar a}$ with $a$ being the lattice spacing and $N$ is the lattice dimension. $n_{\alpha}$ are the Fermi factors: $f(\mu - \epsilon_{\alpha})$ and $\epsilon_{\alpha}$ represents the corresponding eigenenergies. $f^{x}_{\alpha \beta} = <\psi_{\alpha}|j_{x}|\psi_{\beta}>$ are the matrix elements of the current operator $j_{x} = it \sum_{i\sigma} (c^{\dagger}_{i+x,\sigma}c_{i,\sigma} - H.c.)$. The dc conductivity is then obtained by averaging the optical conductivity over a small low frequency interval $\Delta \omega$ - \\

\begin{eqnarray}
\sigma_{av}(\Delta \omega) = \frac{1}{\Delta \omega} \int^{\Delta \omega}_{0} \sigma(\omega) d \omega     \nonumber
\end{eqnarray}

The dc longitudinal conductivity can also be mapped to 

\begin{eqnarray}
\sigma_{xx} = \frac{A}{N} \sum_{\alpha, \beta} (n_{\alpha} - n_{\beta})\frac{|f^{x}_{\alpha \beta}|^{2}}{\epsilon_{\beta} - \epsilon_{\alpha}} \frac{\eta}{(\epsilon_{\beta} - \epsilon_{\alpha})^{2}+\eta^{2}} \nonumber
\end{eqnarray}

similarly, the dc transverse conductivity can be calculated as - 

\begin{eqnarray}
\sigma_{xy} = \frac{A}{N} \sum_{\alpha, \beta} (n_{\alpha} - n_{\beta}) \frac{Im(f^{x}_{\alpha \beta}f^{y}_{\beta\alpha})}{(\epsilon_{\beta} - \epsilon_{\alpha})^{2}+\eta^{2}} \nonumber
\end{eqnarray}

\section{Longitudinal conductivity}
\begin{figure}[h!]
	\centering
	\includegraphics[width=0.45\linewidth]{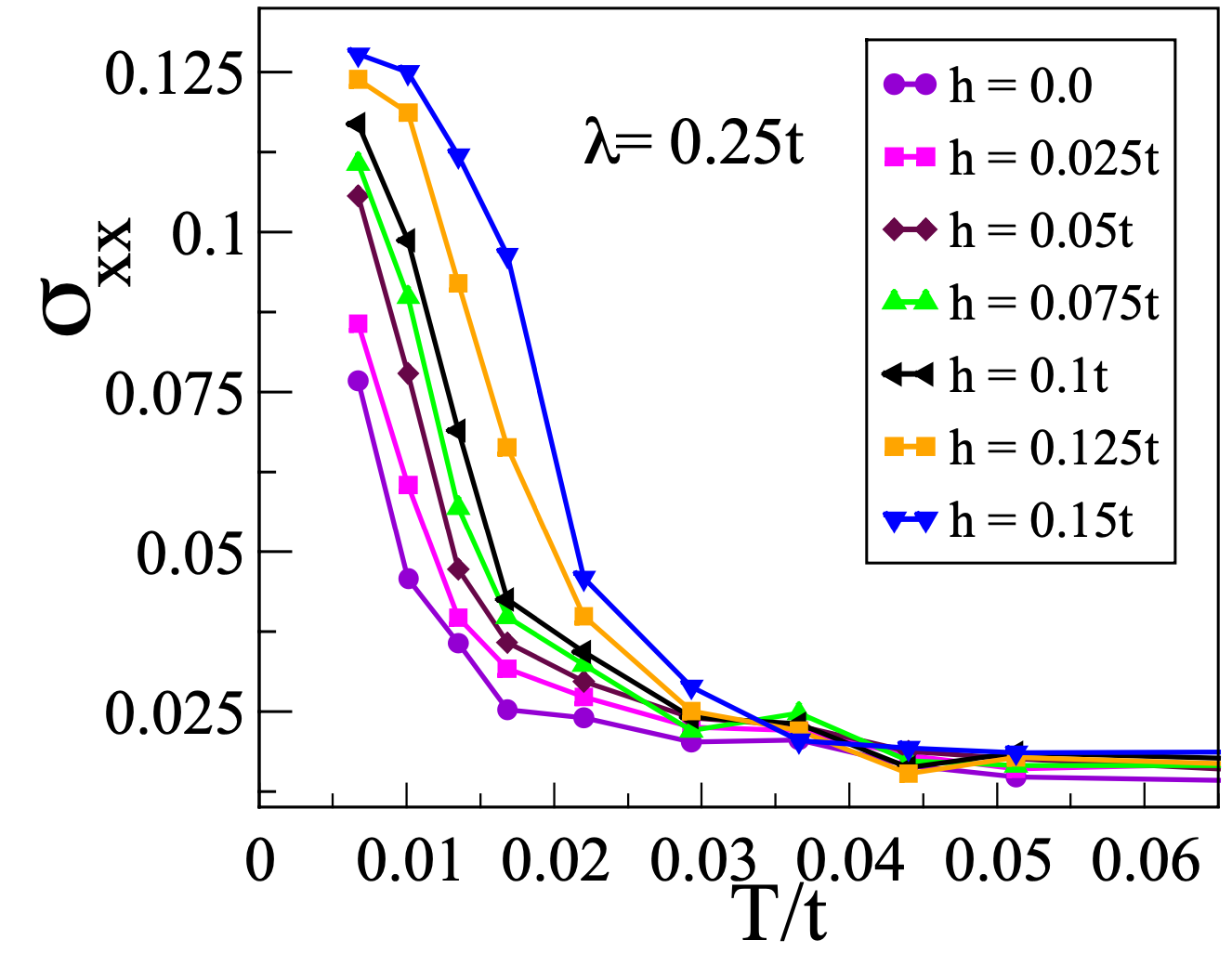}
\caption{ The magnetic field and temperature  dependence of the longitudinal conductivity }
\label{fig:S4}
\end{figure}
We show in Fig.~\ref{fig:S4}, the dependence of $\sigma_{xx}(T)$ on magnetic field. The plots show that while the longitudinal field enhances the magnitude of the longitudinal conductivity by small amounts, $d\sigma_{xx}/dT<0$ for all field values.
\bibliography{sup_ref}